\pdfoutput=1 
\documentclass[10pt,oneside,reqno,english]{amsart}
\usepackage[T1]{fontenc}
\usepackage[latin9]{inputenc}
\usepackage[a4paper]{geometry}
\usepackage{textcomp}
\usepackage{amstext}
\usepackage{amsthm}
\usepackage{amssymb}

\makeatletter

\numberwithin{equation}{section}
\numberwithin{figure}{section}

\usepackage{etoolbox}
\usepackage[hidelinks]{hyperref}
\makeatletter
\patchcmd{\@maketitle}
  {\ifx\@empty\@dedicatory}
  {\ifx\@empty\@date \else {\vskip3ex \centering\footnotesize\@date\par\vskip1ex}\fi
   \ifx\@empty\@dedicatory}
  {}{}
\patchcmd{\@adminfootnotes}
  {\ifx\@empty\@date\else \@footnotetext{\@setdate}\fi}
  {}{}{}
\makeatother

\makeatother

\usepackage{babel}
\begin{document}
\title{Quantum mechanics without quantum potentials}
\author{Adam Brownstein$^{*}$}
\thanks{$^{*}$Melbourne, Australia. ORCID: https://orcid.org/0009-0001-7814-4384\\
}
\date{January 8, 2024}
\begin{abstract}
\noindent The issue of non-locality in quantum mechanics can potentially
be resolved by considering relativistically covariant diffusion in
four-dimensional spacetime. Stochastic particles described by the
Klein-Gordon equation are shown to undergo a classical diffusion process
in spacetime coordinates, which is seen by transforming the quantum
Cauchy-momentum equations to a Lagrangian frame of reference. Since
the quantum potential term is removed under this transformation, the
equations for momentum propagation along particle trajectories assume
a classical form. A local stochastic de Broglie-Bohm interpretation
for the Klein-Gordon system can subsequently be derived. We also introduce
the concept of momentum equivariance to replace the second-order Bohm-Newton
equations of motion, which break down due to non-linear terms of the
stochastic Lagrangian derivative. 
\end{abstract}

\maketitle

\section{Introduction and summary}

\subsection{Introduction }

\noindent In 1966, Nelson \cite{key-1} published a stochastic interpretation
of quantum mechanics which attempted to remove the quantum potential
term and describe quantum mechanics in terms of counter-propagating
classical diffusion processes. The interpretation demonstrated a mysterious
connection between classical diffusion, quantum mechanics, and the
Hamilton-Jacobi formalism. Papers by Bohm, Vigier and Hiley \cite{key-2,key-8}
provided further insights in this direction, where it was shown that
de Broglie-Bohm particles are able to follow a stochastic version
of the guidance equation and still reproduce the quantum predictions. 

Given that the stochastic guidance equations depend upon the transformed
phase $S'=S+\frac{\hbar}{2}\log\rho$, we considered it desirable
to also transform the quantum Hamilton-Jacobi equations so that the
transformed phase can be used to update the particle velocities, instead
of the ordinary quantum phase $S$. In connection to this problem,
we discovered that the quantum potential can be written in a decomposed
form as: 
\begin{equation}
Q=\frac{\hbar^{2}}{8m}\nabla\log\rho\cdot\nabla\log\rho-\frac{\hbar^{2}}{4m}\frac{\nabla^{2}\rho}{\rho},
\end{equation}
which results in a symmetry between the $\nabla\log\rho\cdot\nabla\log\rho$
and $\nabla S\cdot\nabla S$ terms when substituted into the quantum
Hamilton-Jacobi equation. This symmetry is suggestive that the quantum
potential can absorbed into the kinetic energy term using a suitable
transformation. Unfortunately the transformation of the phase $S'=S+\frac{\hbar}{2}\log\rho$,
which we denote as the stochastic transformation, causes the Hamilton-Jacobi
equation to become: 
\begin{equation}
\frac{\partial S'}{\partial t}+\frac{1}{2m}\nabla S'\cdot\nabla S'+\frac{\hbar}{2m}\nabla^{2}S'+V+2Q=0,
\end{equation}
and so far from removing the quantum potential term, the stochastic
transformation doubles it. This is interesting in its own right, nevertheless
it does not answer the question of how to remove the quantum potential
from the equation. After searching for a solution to this problem,
we discovered that the quantum potential has a second decomposition
of the form:
\begin{equation}
\rho\partial_{j}Q=-\frac{\hbar^{2}}{4m}\nabla^{2}\partial_{j}\rho+\frac{\hbar^{2}}{4m}\nabla\cdot\left(\rho\partial_{j}\log\rho\nabla\log\rho\right),
\end{equation}
which is suggestive of applying the stochastic transformation to the
Cauchy-momentum equations, which depend upon $\rho\partial_{j}Q$.
Doing so makes significant progress toward eliminating the quantum
potential, however a complication is that it is necessary to use the
Cauchy-momentum equation as derived from the Klein-Gordon equation
instead of from the non-relativistic Schrödinger equation. After performing
the stochastic transformation upon the relativistic quantum Cauchy-momentum
equations, they become a classical-like Fokker-Planck equation for
momentum conservation: 
\begin{equation}
\frac{\partial\rho\partial_{\nu}S'}{\partial\tau}+\frac{1}{m}\partial_{\mu}\left(\rho\partial_{\nu}S'\partial^{\mu}S'\right)+\partial_{\nu}V=\frac{\hbar}{2m}\partial_{\mu}\partial^{\mu}\left(\rho\partial_{\nu}S'\right).\label{eq:3.22-1-2-1-1}
\end{equation}
\noindent The quantum potential term $\rho\partial_{\nu}Q$ is clearly
absent from this equation, and the additional term $\frac{\hbar}{2m}\partial_{\mu}\partial^{\mu}\left(\rho\partial_{\nu}S'\right)$
can be understood as the non-linear component of a stochastic Lagrangian
derivative. This equation can be written in terms of the stochastic
Lagrangian derivative as follows:
\begin{equation}
\frac{D\rho\partial_{\nu}S'}{D\tau}+\rho\partial_{\nu}S'\left(\frac{1}{m}\partial_{\mu}\partial^{\mu}S'\right)+\rho\partial_{\nu}V=0,\label{eq:3.22-1-2-1-2}
\end{equation}
which is identical in form to the classical Cauchy-momentum equations
in the deterministic Lagrangian reference frame, however it is expressed
in the Lagrangian reference frame of stochastic particles. It is quite
remarkable that the quantum potential can be eliminated from the quantum
Cauchy-momentum equations in this manner. This seems to indicate a
deep importance regarding momentum conservation in understanding quantum
mechanics. These equations potentially provide a path toward a local-realist
interpretation of quantum entanglement. 

\subsection{Executive summary}

\subsubsection{Stochastic transformation \& quantum potential }

A stochastic de Broglie-Bohm interpretation is developed for the Klein-Gordon
system, based upon the \emph{stochastic transformation} $S'=S+\frac{\hbar}{2}\log\rho$.
The stochastic transformation is applied to the continuity equation
and quantum Cauchy-momentum equations respectively. It is shown that
under the stochastic transformation that the continuity equation becomes:
\begin{equation}
\frac{D\rho}{D\tau}+\rho\left(\frac{1}{m}\partial_{\mu}\partial^{\mu}S'\right)=0,\label{eq:3.23-1-1-1-1}
\end{equation}
\noindent while the quantum Cauchy-momentum equation becomes:
\begin{equation}
\frac{D\rho\partial_{\nu}S'}{D\tau}+\rho\partial_{\nu}S'\left(\frac{1}{m}\partial_{\mu}\partial^{\mu}S'\right)+\rho\partial_{\nu}V=0,\label{eq:3.22-1-2-1}
\end{equation}
where $\frac{D}{D\tau}=\frac{\partial}{\partial\tau}+\partial_{\mu}S'\partial^{\mu}-\frac{\hbar}{2m}\partial_{\mu}\partial^{\mu}$
is the forward-causal stochastic Lagrangian derivative. Therefore
the quantum potential term $\rho\partial_{\nu}Q$ is removed by the
stochastic transformation. 

\subsubsection{Rank-2 current conservation}

It is found that the conservation of a rank-2 current density $\rho\partial_{\mu}\partial_{\nu}S$
is required for removing the quantum potential term from the Cauchy-momentum
equations. Conservation of this quantity can be derived for the Klein-Gordon
equation but not the non-relativistic Schrödinger equation. This strongly
suggests the non-conservation of this quantity is related to non-locality. 

\subsubsection{Momentum conservation \& momentum equivariance}

It is shown that the Newtonian second-order form of the dynamics ${\bf a}=-\frac{1}{m}\nabla U$
breaks down in the stochastic case, due to non-linear terms of the
Itô calculus causing $\frac{D\rho\partial_{\nu}S}{D\tau}\neq\rho\frac{D\partial_{\nu}S}{D\tau}+\frac{D\rho}{D\tau}\partial_{\nu}S$.
To resolve this problem, it is demonstrated that point-like de Broglie-Bohm
particles satisfy an equivariance relationship for their momenta.
By this momentum equivariance relationship, it is ensured that particles
undergoing the classical-like Newtonian equations of motion satisfy
the quantum predictions for their momenta. The momentum equivariance
relationship determines the required velocity update law.

Therefore we make an important clarification to the Bohm-Newton second-order
dynamics; that momentum equivariance and the Cauchy-momentum equation
should be used in preference to the gradient of the Hamilton-Jacobi
equations for specifying the equations of motion, since momentum conservation
generalizes the second-order dynamical picture to the stochastic case. 

\subsubsection{Philosophical implications of time}

We explore several philosophical implications of the spacetime aspects
of the interpretation. We suggest that dynamical relaxation to quantum
equilibrium in spacetime coordinates avoids the idea of a static Block
universe. Furthermore we discuss that the interpretation is retrocausal
but not fine-tuned or superdeterminstic. 

We also propose that the local-time coordinate can be understood as
an internal degree of freedom located at the three-dimensional space
positions. In this understanding, the local-time coordinate merely
represents a multivaluedness similar to the wavefunction phase, which
avoids the concept of a physically real four-dimensional spacetime.

\subsubsection{Ancillary results. }

The appendix contains several ancillary results which may be useful
in other contexts. One such result is a derivation of the equivariance
relationship using a phase space approach. Those who are unfamiliar
with the de Broglie-Bohm interpretation may wish to learn about the
equivariance relationship, as it holds the key to understanding the
de Broglie-Bohm interpretation and how it links to the results of
the Copenhagen interpretation. Another result is the derivation of
alternative forms of the quantum potential used within the paper. 

\section{\label{sec:Stochastic-mechanics}Stochastic mechanics}

\subsection{de Broglie-Bohm interpretation}

The de Broglie-Bohm interpretation has two main equations, which are
derived from the real and imaginary components of the Schrödinger
equation. The first equation is the quantum Hamilton-Jacobi equation,
which describes the time evolution of the phase:
\begin{equation}
\frac{\partial S}{\partial t}+\frac{1}{2m}\nabla S\cdot\nabla S+V+Q=0,\label{eq:1-2}
\end{equation}
where $V$ is a classical potential and the quantum potential $Q$
is defined as: 
\begin{equation}
Q\equiv-\frac{\hbar^{2}}{2m}\frac{\nabla^{2}\sqrt{\rho}}{\sqrt{\rho}}.
\end{equation}
The second equation is the continuity equation:
\begin{equation}
\frac{\partial\rho}{\partial t}+\nabla\cdot\left(\rho\frac{\nabla S}{m}\right)=0,\label{eq:2-2}
\end{equation}
which describes the time evolution of the probability density $\rho$.
The logic of the de Broglie-Bohm interpretation is that particles
have trajectories given by the guidance equation: 
\begin{equation}
\frac{d{\bf X}_{t}}{dt}=\frac{\nabla S({\bf x},t)|_{{\bf x}={\bf X}_{t}}}{m},\label{eq:3}
\end{equation}
where the conditioning symbol $|_{{\bf x}={\bf X}_{t}}$ indicates
that the function is evaluated at the actual location of the particle
configuration ${\bf X}_{t}$. A statistical ensemble of de Broglie-Bohm
particles which follow this guidance equation will reproduce the quantum
probability distribution $\rho$ due to the equivariance property.

\subsection{Stochastic de Broglie-Bohm interpretation}

It was recognized by Bohm, Vigier \& Hiley \cite{key-2,key-8} that
de Broglie-Bohm particles can have stochastic trajectories and still
be consistent with the continuity equation of quantum mechanics. To
see this, add $\frac{\hbar}{2m}\nabla^{2}\rho$ to both sides of the
continuity equation:
\begin{equation}
\frac{\partial\rho}{\partial t}+{\bf \nabla}\cdot\left[\rho\frac{{\bf {\bf \nabla}}\left(S+\frac{\hbar}{2}\log\rho\right)}{m}\right]=\frac{\hbar}{2m}{\bf \nabla}^{2}\rho,\label{eq:2.5}
\end{equation}
where the relation $\frac{\nabla\rho}{\rho}=\nabla\log\rho$ has been
used. This equation can be simplified by defining a transformed phase
$S'=S+\frac{\hbar}{2}\log\rho$ which gives:
\begin{equation}
\frac{\partial\rho}{\partial t}+{\bf \nabla}\cdot\left(\rho\frac{{\bf \nabla}S'}{m}\right)=\frac{\hbar}{2m}{\bf \nabla}^{2}\rho.
\end{equation}
The transformed continuity equation is a Fokker-Planck equation which
describes a statistical ensemble of particles undergoing motion governed
by the stochastic guidance equation:
\begin{equation}
d{\bf X}_{t}=\frac{1}{m}\nabla S'({\bf x},t)|_{{\bf x}={\bf X}_{t}}dt+\sqrt{\frac{\hbar}{m}}d{\bf W}_{t},\label{eq:13-1}
\end{equation}
where $d{\bf W}_{t}$ is the Wiener process of Brownian motion.

\subsection{Klein-Gordon equation: de Broglie-Bohm interpretation}

Now we will derive the equations of the deterministic de Broglie-Bohm
interpretation for the Klein-Gordon system. Using the metric signature
$\eta=(-1,1,1,1)$, the Klein-Gordon equation can be written as: 
\begin{equation}
\left[\partial_{\mu}\partial^{\mu}-\frac{m^{2}c^{2}}{\hbar^{2}}-\frac{2m}{\hbar^{2}}V\right]\psi=0,\label{eq:1-3-1}
\end{equation}
where we have rescaled the classical potential $V$ by a factor of
$\left(\frac{2m}{\hbar^{2}}\right)^{-1}$. The wavefunction $\psi$
for the Klein-Gordon equation is a complex valued scalar, and has
a polar form of:
\begin{equation}
\psi=\sqrt{\rho}e^{i\frac{S}{\hbar}}.
\end{equation}
Substitute the polar form of the wavefunction into {[}Eq. \ref{eq:1-3-1}{]}
and separate the real and imaginary components. This is completed
in Appendix \ref{sec:Polar-decomposition-of}. The two resulting equations
are firstly a relativistic generalization of the continuity equation:
\begin{equation}
\partial_{\mu}\left(\rho\frac{\partial^{\mu}S}{m}\right)=0,\label{eq:209}
\end{equation}
and secondly a relativistic generalization of the quantum Hamilton-Jacobi
equation:
\begin{equation}
\frac{1}{2m}\partial_{\mu}S\partial^{\mu}S+V+Q+\frac{1}{2}mc^{2}=0,\label{eq:2-10}
\end{equation}
where the relativistic quantum potential is defined as:
\begin{equation}
Q\equiv-\frac{\hbar^{2}}{2m}\frac{\partial_{\mu}\partial^{\mu}\sqrt{\rho}}{\sqrt{\rho}}.
\end{equation}
Since solutions to the Klein-Gordon equation do not depend on proper
time $\tau$, we also have the additional constraints:
\begin{eqnarray}
\frac{\partial\rho}{\partial\tau}=0, &  & \frac{\partial S}{\partial\tau}=0.
\end{eqnarray}
These constraints can be combined with {[}Eq. \ref{eq:209}{]} and
{[}Eq. \ref{eq:2-10}{]} to write the continuity equation and quantum
Hamilton-Jacobi equation in a more similar form to the non-relativistic
case:
\begin{equation}
\frac{\partial\rho}{\partial\tau}+\partial_{\mu}\left(\rho\frac{\partial^{\mu}S}{m}\right)=0,\label{eq:215}
\end{equation}

\begin{equation}
\frac{\partial S}{\partial\tau}+\frac{1}{2m}\partial_{\mu}S\partial^{\mu}S+V+Q+\frac{1}{2}mc^{2}=0.
\end{equation}
A de Broglie-Bohm interpretation of the Klein-Gordon system can now
be obtained from these equations. Assume de Broglie-Bohm particles
are located with position $X_{\tau}^{\mu}$ in a configuration space
of spacetime coordinates. We then define the de Broglie-Bohm guidance
equation to be:
\begin{equation}
\frac{dX_{\tau}^{\mu}}{d\tau}=\frac{1}{m}\partial^{\mu}S(x)|_{x=X_{\tau}}.\label{eq:24-1}
\end{equation}
A statistical ensemble of particle configurations with this guidance
equation will reproduce the probability density $\rho(x,\tau)=\rho(x)$
for all proper times $\tau$ due to the equivariance property. Moreover,
if the particle ensemble is prepared with a distribution different
to $\rho(x)$, it will dynamically relax to this quantum equilibrium
distribution. Refer to Valentini \cite{key-3} for a discussion of
dynamical relaxation to quantum equilibrium. 

\subsection{A note on proper-time independence}

The proper time $\tau$ is a background coordinate which is not present
in the Klein-Gordon equations. Therefore, it has no bearing on the
dynamical system at the level of wavefunctions, which are unaware
of proper time. The partial derivatives of quantities with respect
to proper time such as $\frac{\partial\rho}{\partial\tau}$ and $\frac{\partial S}{\partial\tau}$
are therefore known to vanish. Adding these terms to the continuity
equation and quantum Hamilton-Jacobi equation therefore has no effect,
as any valid solution to the equations already satisfy these constraints. 

\subsection{\label{subsec:Klein-Gordon-equation:-stochasti}Klein-Gordon equation:
stochastic de Broglie-Bohm interpretation}

Similar to the non-relativistic case, the de Broglie-Bohm interpretation
of the Klein-Gordon system has a stochastic version. By adding $\frac{\hbar}{2m}\partial_{\mu}\partial^{\mu}\rho$
to both sides of the continuity equation {[}Eq. \ref{eq:215}{]} we
get:
\begin{equation}
\frac{\partial\rho}{\partial\tau}+\partial_{\mu}\left[\rho\partial^{\mu}\frac{\left(S+\frac{\hbar}{2}\log\rho\right)}{m}\right]=\frac{\hbar}{2m}\partial_{\mu}\partial^{\mu}\rho.
\end{equation}
Then by using the transformed phase $S'=S+\frac{\hbar}{2}\log\rho$,
the continuity equation becomes a Fokker-Planck equation:
\begin{equation}
\frac{\partial\rho}{\partial\tau}+\partial_{\mu}\left(\rho\frac{\partial^{\mu}S'}{m}\right)=\frac{\hbar}{2m}\partial_{\mu}\partial^{\mu}\rho.\label{eq:2.18}
\end{equation}
Therefore we propose a stochastic guidance equation of the form:
\begin{equation}
\text{d}X_{\tau}^{\mu}=\frac{1}{m}\partial^{\mu}S'(x)|_{x=X_{\tau}}+\sqrt{\frac{\hbar}{m}}\text{dW}^{\mu},\label{eq:26}
\end{equation}
where $\text{dW}^{\mu}$ is a Wiener process. A statistical ensemble
of particles with this guidance equation, if prepared with distribution
equal to the quantum mechanical probability density $\rho(x,\tau_{0})=\rho(x)$,
are guaranteed to reproduce this distribution for all subsequent proper
times $\tau$ due to the equivariance property. If the particle ensemble
is prepared with a distribution different to $\rho(x)$, it will dynamically
relax to this quantum equilibrium distribution \cite{key-3}. The
continuity equation {[}Eq. \ref{eq:2.18}{]} can also be written in
terms of the forward-causal stochastic Lagrangian derivative as:
\begin{equation}
\frac{D\rho}{D\tau}+\rho\left(\frac{1}{m}\partial_{\mu}\partial^{\mu}S'\right)=0,\label{eq:3.23-1}
\end{equation}
where the forward-causal stochastic Lagrangian derivative is defined
as: 

\begin{equation}
\frac{D}{D\tau}\equiv\frac{\partial}{\partial\tau}+\frac{1}{m}\partial_{\mu}S'\partial_{\mu}-\frac{\hbar}{2m}\partial_{\mu}\partial^{\mu}.
\end{equation}

\subsection{Many-body notation}

For simplicity of notation, we adopt Albert's configuration space
point ontology \cite{key-5}. The idea is that the equations for a
particle configuration are the same as those for a single particle
when the particle configuration is considered as a single point in
configuration space. This allows us to write the equations for a many-body
system in a simplified form. For example, in a bipartite system, instead
of writing the continuity equation explicitly in terms of both particles
as:
\begin{equation}
\frac{\partial\rho(x_{1},x_{2},t)}{\partial t}+\nabla_{1}\cdot\left(\rho(x_{1},x_{2},t)\frac{\nabla_{1}S(x_{1},x_{2},t)}{m}\right)+\nabla_{2}\cdot\left(\rho(x_{1},x_{2},t)\frac{\nabla_{2}S(x_{1},x_{2},t)}{m}\right)=0,\label{eq:2-2-1}
\end{equation}
it can be written in configuration space point notation as: 
\begin{equation}
\frac{\partial\rho(x,t)}{\partial t}+\nabla\cdot\left(\rho(x,t)\frac{\nabla S(x,t)}{m}\right)=0,
\end{equation}
where $x=(x_{1},x_{2})$ is an element of the configuration space
$\left(\mathbb{R}^{3}\right)^{\otimes2}\equiv\mathbb{R}^{3}\otimes\mathbb{R}^{3}$.
In the relativistic case, the particle positions are coordinates in
four-dimensional spacetime. Therefore the particle configuration is
written as $x=(x_{1},...,x_{N})$, where $x$ is an element of the
configuration space $\left(\mathbb{R}\otimes\mathbb{R}^{3}\right)^{\otimes N}$.
For example, the relativistic continuity equation in the two-particle
case can be written explicitly as: 
\begin{equation}
\frac{\partial\rho(x_{1},x_{2},\tau)}{\partial\tau}+\partial_{\mu_{1}}\left(\rho(x_{1},x_{2},\tau)\partial^{\mu_{1}}\frac{S(x_{1},x_{2},\tau)}{m}\right)+\partial_{\mu_{2}}\left(\rho(x_{1},x_{2},\tau)\partial^{\mu_{2}}\frac{S(x_{1},x_{2},\tau)}{m}\right)=0,
\end{equation}
or more simply in the configuration space point notation as: 
\begin{equation}
\frac{\partial\rho(x,\tau)}{\partial\tau}+\partial_{\mu}\left(\rho(x,\tau)\frac{\partial^{\mu}S(x,\tau)}{m}\right)=0.
\end{equation}
Using this simplified notation, the equations can be understood for
either the single particle or particle configuration as desired. However,
it should be noted that the results of this paper are deeply concerned
with the issue of non-locality, and the configuration space interpretation
is intended. 

In the non-relativistic case, this is a single-time formalism for
the many-body system, while in the relativistic case, it is a multi-time
formalism. While there is only a single proper time in the relativistic
case, the local-time coordinates should be considered on the same
footing as the spatial coordinates as elements of the configuration
space. In the relativistic case then, it evident from the multi-time
formalism that entanglement correlations can occur in the local-time
degrees of freedom as well the spatial degrees of freedom. 

\section{Cauchy-momentum equations\label{sec:Cauchy-momentum-equations}}

\subsection{Overview }

In this section we show that the quantum potential can be removed
by absorbing it into the kinetic terms of the quantum Cauchy-momentum
equation, using the stochastic transformation $S'=S+\frac{\hbar}{2}\log\rho$.
To prove this result, we will firstly derive the quantum Cauchy-momentum
equations for the Klein-Gordon system. Then the stochastic transform
is applied, whereby it is shown that the quantum potential term disappears.
In Appendix \ref{subsec:Non-relativistic-case} we also investigate
whether the same result holds true for the non-relativistic case of
the quantum Hamilton-Jacobi equations.

\subsection{Klein-Gordon system}

The quantum Cauchy-momentum equations can be derived from the relativistic
quantum Hamilton-Jacobi and continuity equations as follows. Firstly,
take the derivative of the quantum Hamilton-Jacobi equation {[}Eq.
\ref{eq:2-10}{]}, which gives:
\begin{align}
0= & \frac{1}{2m}\partial_{\nu}\left(\partial_{\mu}S\partial^{\mu}S\right)+\partial_{\nu}V+\partial_{\nu}Q\\
= & \frac{1}{m}\partial_{\mu}S\partial^{\mu}\partial_{\nu}S+\partial_{\nu}V+\partial_{\nu}Q.
\end{align}
Now multiply through by $\rho$:
\begin{equation}
\frac{1}{m}\rho\partial_{\mu}S\partial^{\mu}\partial_{\nu}S+\rho\partial_{\nu}V+\rho\partial_{\nu}Q=0.
\end{equation}
Factorize the derivative operator $\partial^{\mu}$ from the first
term using the inverse product rule:
\begin{equation}
\frac{1}{m}\partial_{\mu}\left(\rho\partial_{\nu}S\partial^{\mu}S\right)-\frac{1}{m}\partial_{\mu}\left(\rho\partial^{\mu}S\right)\partial_{\nu}S+\rho\partial_{\nu}V+\partial_{\nu}Q=0.
\end{equation}
By the continuity equation {[}Eq. \ref{eq:209}{]} we have $\partial_{\mu}\left(\rho\partial^{\mu}S\right)=0$,
which leaves:
\begin{equation}
\frac{1}{m}\partial_{\mu}\left(\rho\partial_{\nu}S\partial^{\mu}S\right)+\rho\partial_{\nu}V+\rho\partial_{\nu}Q=0.\label{eq:53}
\end{equation}
This is the quantum Cauchy-momentum equation. We can make the interpretation
of this equation more evident by adding $\frac{\partial\rho\partial_{\nu}S}{\partial\tau}=0$,
which follows since $\rho(x)$ or $\partial_{\nu}S(x)$ are assumed
not to depend on $\tau$. Therefore the quantum Cauchy-momentum equation
can be written as:
\begin{equation}
\frac{\partial\rho\partial_{\nu}S}{\partial\tau}+\frac{1}{m}\partial_{\mu}\left(\rho\partial_{\nu}S\partial^{\mu}S\right)+\rho\partial_{\nu}V+\rho\partial_{\nu}Q=0.
\end{equation}
This equation is of the same form as a continuity equation but with
$\rho$ replaced with $\rho\partial_{\nu}S$, and the appearance of
source terms for the momenta $\rho\partial_{\nu}V$ and $\rho\partial_{\nu}Q$,
which indicate it is an equation describing momentum conservation.

\subsection{\label{subsec:Stochastic-transformation}Stochastic transformation }

The quantum Cauchy-momentum equations can undergo a stochastic transformation
$S'=S+\frac{\hbar}{2}\log\rho$ similar to that performed in section
\ref{subsec:Klein-Gordon-equation:-stochasti} for the continuity
equation. It is shown in Appendix \ref{sec:Quantum-potential} that
$\rho\partial_{\nu}Q$ can be decomposed into the following useful
form {[}Eq. \ref{eq: D11}{]}:
\begin{equation}
\rho\partial_{\nu}Q=-\frac{\hbar^{2}}{4m}\partial_{\mu}\partial^{\mu}\partial_{\nu}\rho+\frac{\hbar^{2}}{4m}\partial_{\mu}\left(\rho\partial_{\nu}\log\rho\partial^{\mu}\log\rho\right).
\end{equation}
Substitute this form of the quantum potential into the Cauchy-momentum
equation {[}Eq. \ref{eq:53}{]} to give:
\begin{equation}
\frac{1}{m}\partial_{\mu}\left(\rho\partial_{\nu}S\partial^{\mu}S\right)+\frac{\hbar^{2}}{4m}\partial_{\mu}\left(\rho\partial_{\nu}\log\rho\partial^{\mu}\log\rho\right)-\frac{\hbar^{2}}{4m}\partial_{\mu}\partial^{\mu}\partial_{\nu}\rho+\rho\partial_{\nu}V=0.
\end{equation}
Collect like terms:
\begin{equation}
\frac{1}{m}\partial_{\mu}\left[\rho\left(\partial_{\nu}S\partial^{\mu}S+\frac{\hbar^{2}}{4m}\partial_{\nu}\log\rho\partial^{\mu}\log\rho\right)\right]+\rho\partial_{\nu}V=\frac{\hbar^{2}}{4m}\partial_{\mu}\partial^{\mu}\partial_{\nu}\rho.
\end{equation}
This equation is suggestive of combining the $\partial_{\nu}S\partial^{\mu}S$
term with the $\frac{\hbar^{2}}{4m}\partial_{\nu}\log\rho\partial^{\mu}\log\rho$
term using the stochastic transformation $S'=S+\frac{\hbar}{2}\log\rho$.
To do this, first add and subtract the necessary cross terms $\partial_{\nu}S\partial^{\mu}\log\rho$
and $\partial_{\nu}\log\rho\partial^{\mu}S$:
\begin{align}
\frac{1}{m}\partial_{\mu}\left[\rho\left(\partial_{\nu}S\partial^{\mu}S+\frac{\hbar}{2}\partial_{\nu}S\partial^{\mu}\log\rho+\frac{\hbar}{2}\partial_{\nu}\log\rho\partial^{\mu}S+\frac{\hbar^{2}}{4}\partial_{\nu}\log\rho\partial^{\mu}\log\rho\right)\right]\ldots\nonumber \\
\ldots-\frac{\hbar}{2m}\partial_{\mu}\left(\rho\partial_{\nu}S\partial^{\mu}\log\rho\right)-\frac{\hbar}{2m}\partial_{\mu}\left(\rho\partial_{\nu}\log\rho\partial^{\mu}S\right)+\rho\partial_{\nu}V=\frac{\hbar^{2}}{4m}\partial_{\mu}\partial^{\mu}\partial_{\nu}\rho,
\end{align}
which simplifies to:
\begin{equation}
\frac{1}{m}\partial_{\mu}\left(\rho\partial_{\nu}S'\partial^{\mu}S'\right)-\frac{\hbar}{2m}\partial_{\mu}\left(\partial_{\nu}S\partial^{\mu}\rho\right)-\frac{\hbar}{2m}\partial_{\mu}\left(\partial_{\nu}\rho\partial^{\mu}S\right)+\rho\partial_{\nu}V=\frac{\hbar^{2}}{4m}\partial_{\mu}\partial^{\mu}\partial_{\nu}\rho.\label{eq:3.10}
\end{equation}
Now we examine the cross terms individually in order to simplify further.
The first cross term is equal to:
\begin{align}
-\frac{\hbar}{2m}\partial_{\mu}\left(\partial_{\nu}S\partial^{\mu}\rho\right)= & -\frac{\hbar}{2m}\partial_{\mu}\partial^{\mu}\left(\rho\partial_{\nu}S\right)+\frac{\hbar}{2m}\partial_{\mu}\left(\rho\partial^{\mu}\partial_{\nu}S\right).\label{eq:3.11}
\end{align}
The second cross term is equal to: 
\begin{align}
-\frac{\hbar}{2m}\partial_{\mu}\left(\partial_{\nu}\rho\partial^{\mu}S\right)= & -\frac{\hbar}{2m}\partial_{\nu}\partial_{\mu}\left(\rho\partial^{\mu}S\right)+\frac{\hbar}{2m}\partial_{\mu}\left(\rho\partial^{\mu}\partial_{\nu}S\right)\\
= & \frac{\hbar}{2m}\partial_{\mu}\left(\rho\partial^{\mu}\partial_{\nu}S\right),\label{eq:3.13}
\end{align}
which follows since $\partial_{\mu}\left(\rho\partial^{\mu}S\right)=0$
by the continuity equation {[}Eq. \ref{eq:209}{]}. Therefore substituting
these expressions ({[}Eq. \ref{eq:3.11}{]}, {[}Eq. \ref{eq:3.13}{]})
into {[}Eq. \ref{eq:3.10}{]} gives:
\begin{equation}
\frac{1}{m}\partial_{\nu}\left(\rho\partial_{\mu}S'\partial^{\nu}S'\right)+\frac{\hbar}{m}\partial_{\mu}\left(\rho\partial^{\mu}\partial_{\nu}S\right)+\rho\partial_{\nu}V=\frac{\hbar^{2}}{4m}\partial_{\mu}\partial^{\mu}\partial_{\nu}\rho+\frac{\hbar}{2m}\partial_{\mu}\partial^{\mu}\left(\rho\partial_{\nu}S\right).\label{eq:3.15}
\end{equation}
On the right-hand-side, the two terms can be combined using the stochastic
transform:
\begin{align}
\frac{\hbar^{2}}{4m}\partial_{\mu}\partial^{\mu}\partial_{\nu}\rho+\frac{\hbar}{2m}\partial_{\mu}\partial^{\mu}\left(\rho\partial_{\nu}S\right)= & \frac{\hbar}{2m}\partial_{\mu}\partial^{\mu}\left(\rho(\partial_{\nu}S+\frac{\hbar}{2}\log\rho)\right)\\
= & \frac{\hbar}{2m}\partial_{\mu}\partial^{\mu}\left(\rho\partial_{\nu}S'\right),\label{eq:3.16}
\end{align}
therefore using {[}Eq. \ref{eq:3.16}{]}, {[}Eq. \ref{eq:3.15}{]}
becomes: 
\begin{equation}
\frac{1}{m}\partial_{\nu}\left(\rho\partial_{\mu}S'\partial^{\nu}S'\right)+\frac{\hbar}{m}\partial_{\mu}\left(\rho\partial^{\mu}\partial_{\nu}S\right)+\rho\partial_{\nu}V=\frac{\hbar}{2m}\partial_{\mu}\partial^{\mu}\left(\rho\partial_{\nu}S'\right).\label{eq:3.17}
\end{equation}
For the term $\partial_{\mu}\left(\rho\partial^{\mu}\partial_{\nu}S\right)$,
we assume that this term equals zero, which will be demonstrated in
section \ref{sec:Continuity-condition}. Setting $\partial_{\mu}\left(\rho\partial^{\mu}\partial_{\nu}S\right)=0$
in {[}Eq. \ref{eq:3.17}{]} then gives: 
\begin{equation}
\frac{1}{m}\partial_{\nu}\left(\rho\partial_{\mu}S'\partial^{\nu}S'\right)+\rho\partial_{\nu}V=\frac{\hbar}{2m}\partial_{\mu}\partial^{\mu}\left(\rho\partial_{\nu}S'\right).\label{eq:3.19A}
\end{equation}
Equivalently, by adding $\frac{\partial\rho\partial_{\nu}S'}{\partial\tau}=0$
to this equation:
\begin{equation}
\frac{\partial\rho\partial_{\nu}S'}{\partial\tau}+\frac{1}{m}\partial_{\nu}\left(\rho\partial_{\mu}S'\partial^{\nu}S'\right)+\rho\partial_{\nu}V=\frac{\hbar}{2m}\partial_{\mu}\partial^{\mu}\left(\rho\partial_{\nu}S'\right).\label{eq:3.19}
\end{equation}
This equation {[}Eq. \ref{eq:3.19A}, \ref{eq:3.19}{]} is the quantum
Cauchy-momentum equation under the stochastic transformation. We can
alternatively express the equation in the Lagrangian reference frame
of stochastic particles by writing it in terms of the forward-causal
stochastic Lagrangian derivative $\frac{D}{D\tau}$ defined as: 
\begin{equation}
\frac{D}{D\tau}=\frac{\partial}{\partial\tau}+\frac{1}{m}\partial_{\mu}S'\partial_{\mu}-\frac{\hbar}{2m}\partial_{\mu}\partial^{\mu}.
\end{equation}
To proceed, expand out the kinetic term of {[}Eq. \ref{eq:3.19}{]}:
\begin{equation}
\frac{\partial\rho\partial_{\nu}S}{\partial\tau}+\frac{1}{m}\left(\partial^{\mu}S'\right)\partial_{\mu}\left(\rho\partial_{\nu}S'\right)-\frac{\hbar}{2m}\partial_{\mu}\partial^{\mu}\left(\rho\partial_{\nu}S'\right)+\frac{1}{m}\left(\rho\partial_{\nu}S'\right)\left(\partial_{\mu}\partial^{\mu}S'\right)+\rho\partial_{\nu}V=0,
\end{equation}
therefore: 
\begin{equation}
\frac{D\rho\partial_{\nu}S}{D\tau}+\rho\partial_{\nu}S'\left(\frac{1}{m}\partial_{\mu}\partial^{\mu}S'\right)+\rho\partial_{\nu}V=0.\label{eq:3.22}
\end{equation}
Notably the equation {[}Eq. \ref{eq:3.22}{]} does not contain the
quantum potential term $\rho\partial_{\nu}Q$, which has been removed
by the stochastic transformation. 

\section{\label{sec:Continuity-condition}Conservation condition}

\noindent In this section we wish to demonstrate the condition $\partial_{\mu}\left(\rho\partial_{\nu}\partial^{\mu}S\right)=0$,
which has been used to prove the results of section \ref{subsec:Stochastic-transformation}.
It is known that the four-vector field $\partial^{\mu}S$ is locally
able to undergo a Lorentz transformation to an inertial reference
frame. In the inertial reference frame, it can be shown via the continuity
equation that $\rho$ is a log-separable function of the spacetime
coordinates. Reconciling this fact with the Fourier mode decomposition
of $\rho$ implies that $\partial_{0'}\rho=0$, i.e. the probability
density is stationary in the inertial reference frame. We also discuss
how this condition is compatible with the quantum Hamilton-Jacobi
equation, which provides further evidence for the condition. Given
$\partial_{0'}\rho=0$, it can easily be shown that $\partial_{0'}(\rho\partial_{0'}\partial^{0'}S)=0$
via the continuity equation, which in turn implies $\partial_{\mu}\left(\rho\partial_{\nu}\partial^{\mu}S\right)=0$. 

\subsection{Conservation condition $\partial_{\mu}\left(\rho\partial_{\nu}\partial^{\mu}S\right)=0$
given $\partial_{0'}\rho=0$\label{subsec:Conservation-condition-}}

Note that the four-vector $\partial^{\mu}S$ has a Lorentz transformation
to an inertial reference frame (primed frame) where $\partial'S=(\partial^{0'}S,0,0,0)$.
In this reference frame, the continuity equation becomes: 
\begin{equation}
\partial_{0'}\left(\rho\partial_{0'}S\right)=0.\label{eq:4.1}
\end{equation}
If we assume $\partial_{0'}\rho=0$, the conservation of the rank-2
current density\emph{ $\rho\partial_{\nu}\partial^{\mu}S$ (conservation
condition)} is simple to show. Firstly, for $\nu'=j'\in[1,2,3]$,
this relation is automatically satisfied since $\partial_{j'}S=0$
which results in $\partial_{\mu'}\left(\rho\partial_{j'}\partial^{\mu'}S\right)=0$.
Therefore we concentrate on the remaining case $\nu'=0$. The continuity
equation {[}Eq. \ref{eq:4.1}{]} in expanded form is:
\begin{equation}
\partial_{0'}\rho\partial^{0'}S+\rho\partial_{0'}\partial^{0'}S=0.\label{eq:4.2}
\end{equation}
Since it is assumed $\partial_{0'}\rho=0$, we have:
\begin{equation}
\rho\partial_{0'}\partial^{0'}S=0.
\end{equation}
This implies:
\begin{align}
\partial_{\mu'}\left(\rho\partial_{0'}\partial^{\mu'}S\right)= & \partial_{0'}\left(\rho\partial_{0'}\partial^{0'}S\right)=0,
\end{align}
which follows since $\partial^{\mu'}S$ is zero for $\mu'\in[1,2,3]$.
Therefore putting together the equations $\partial_{\mu'}\left(\rho\partial_{0'}\partial^{\mu'}S\right)=0$
and $\partial_{\mu'}\left(\rho\partial_{j'}\partial^{\mu'}S\right)=0$
gives:
\begin{equation}
\partial_{\mu'}\left(\rho\partial_{\nu'}\partial^{\mu'}S\right)=0.
\end{equation}
Transforming the $\mu'$ index back to the unprimed frame results
in: 
\begin{equation}
\partial_{\mu}\left(\rho\partial_{\nu'}\partial^{\mu}S\right)=0.
\end{equation}
Now the $\nu'$ index can be transformed back to the unprimed frame
as follows:
\begin{align}
\partial_{\mu}\left(\rho\partial_{\nu}\partial^{\mu}S\right)= & \eta_{\nu}^{\,\,\,\nu'}\partial_{\mu}\left(\rho\partial_{\nu'}\partial^{\mu}S\right)=0,
\end{align}
which proves the result. 

\subsection{$\rho$ is log-separable}

Here we demonstrate that $\rho$ is a log-separable function in the
inertial reference frame. By log-separable we mean that $\log\rho(x_{0'},{\bf x}')=g_{0'}(x_{0'})+g_{{\bf x}'}({\bf x}')$,
which is separable into a function of the time coordinate $g(x_{0'})$
and spatial coordinates $g_{{\bf x'}}({\bf x}')$. Dividing the expanded
continuity equation {[}Eq. \ref{eq:4.2}{]} by $\rho$ gives: 
\begin{equation}
\partial_{0'}\log\rho\partial_{0'}S+\partial_{0'}\partial_{0'}S=0.
\end{equation}
Since $\partial_{0'}\partial_{j'}S=0$, taking the derivative with
respect to the spatial coordinates $j'$ gives: 
\begin{equation}
\partial_{0'}\partial_{j'}\log\rho\partial_{0'}S=0.
\end{equation}
Now since $\partial_{0'}S\neq0$ (due to the quantum Hamilton-Jacobi
equation), divide by $\partial_{0'}S$: 
\begin{equation}
\partial_{0'}\partial_{j'}\log\rho=0.
\end{equation}
This implies that $\rho$ is a log-separable function:
\begin{equation}
\log\rho=g_{0'}(x_{0'})+g_{{\bf x}'}({\bf x}'),
\end{equation}
or equivalently: 
\begin{equation}
\rho=e^{g_{0'}(x_{0'})+g_{{\bf x}'}({\bf x}')}.
\end{equation}

\subsection{Fourier mode decomposition shows $\partial_{0'}\rho=0$\label{subsec:Fourier-mode-decomposition}}

In the following discussion, we will switch to the metric signature
$\eta=(1,-1,-1,-1)$, which ensures the time-component of the Fourier
basis is positive in the rest frame. It is known that the Klein-Gordon
field can be decomposed into Fourier modes as follows:
\begin{align}
\psi= & \sum_{n}a_{n}e^{-ik_{n}^{\alpha}x_{\alpha}}+b_{n}e^{ik_{n}^{\alpha}x_{\alpha}},
\end{align}
where $a_{n}$ and $b_{n}$ are complex scalars. The wave-vectors
$k_{n}^{\mu}$ satisfy the on mass-shell condition: 
\begin{equation}
k_{n}^{\mu}k_{n\mu}=m^{2}c^{2}.
\end{equation}
The positive frequency modes correspond to particles and the negative
frequency modes to antiparticles. Since the particle and antiparticle
modes act in different sectors in any physical theory, i.e. they are
part of separate Hilbert spaces, we restrict our attention to the
particle sector and remove the negative frequency modes from the Fourier
decomposition. Therefore, the particle wavefunction has the following
Fourier mode decomposition:
\begin{align}
\psi= & \sum_{n}a_{n}e^{-ik_{n}^{\alpha'}x_{\alpha'}},
\end{align}
where the equation has been Lorentz transformed to the inertial reference
frame where $\partial'S=(\partial^{0'}S,0,0,0)$, as indicated by
the primed Lorentz indices. Inserting the Fourier decomposition into
the definition for $\rho$ gives:
\begin{align}
\rho= & \psi^{\dagger}\psi\\
= & \sum_{n,m}a_{n}^{\dagger}a_{m}e^{-i\left(k_{m}^{\alpha'}-k_{n}^{\alpha'}\right)x_{\alpha'}}.\label{eq:4.14}
\end{align}
It will not be possible to make this function log-separable unless
$k_{m}^{0'}-k_{n}^{0'}=0$ for any non-zero $a_{n},a_{m}$. This result
should be intuitive because $\rho$ is equal to the sum of log-separable
functions, and the sum of log-separable functions is not itself log-separable
in general. However, to demonstrate the result mathematically, the
quantity $\rho^{2}\partial_{0'}\partial_{j'}\log\rho$ can be calculated
using the Fourier mode decomposition, whereby it can be shown that
the integral of this quantity over the function $e^{i\left(k_{a}^{\alpha'}+k_{c}^{\alpha'}-k_{b}^{\alpha'}-k_{d}^{\alpha'}\right)x_{\alpha'}}$
is non-zero, in contradiction to $\rho^{2}\partial_{0'}\partial_{j'}\log\rho=0$
which is implied by the log-separable nature of $\rho$. To proceed,
the quantity $\rho^{2}\partial_{0'}\partial_{j'}\log\rho$ can firstly
be decomposed using the inverse product rule:
\begin{align}
\rho^{2}\partial_{0'}\partial_{j'}\log\rho= & \rho\partial_{0'}\partial_{j'}\rho-\partial_{0'}\rho\partial_{j'}\rho.
\end{align}
Therefore:
\begin{align}
\int\rho^{2}\partial_{0'}\partial_{j'}\log\rho e^{i\left(k_{a}^{\alpha'}+k_{c}^{\alpha'}-k_{b}^{\alpha'}-k_{d}^{\alpha'}\right)x_{\alpha'}}d^{4}x= & \int\rho\partial_{0'}\partial_{j'}\rho e^{i\left(k_{a}^{\alpha'}+k_{c}^{\alpha'}-k_{b}^{\alpha'}-k_{d}^{\alpha'}\right)x_{\alpha'}}d^{4}x\ldots\nonumber \\
 & \ldots-\int\partial_{0'}\rho\partial_{j'}\rho e^{i\left(k_{a}^{\alpha'}+k_{c}^{\alpha'}-k_{b}^{\alpha'}-k_{d}^{\alpha'}\right)x_{\alpha'}}d^{4}x.\label{eq:4.23-1}
\end{align}
Using the Fourier mode decomposition, it can be shown that:
\begin{align}
 & \int\rho\partial_{0'}\partial_{j'}\rho e^{i\left(k_{a}^{\alpha'}+k_{c}^{\alpha'}-k_{b}^{\alpha'}-k_{d}^{\alpha'}\right)x_{\alpha'}}d^{4}x\nonumber \\
 & =\int\sum_{n,m,r,s}-a_{n}^{\dagger}a_{s}^{\dagger}a_{m}a_{r}\left(k_{m}^{0'}-k_{n}^{0'}\right)\left(k_{m}^{j'}-k_{n}^{j'}\right)e^{-i\left(k_{m}^{\alpha'}+k_{r}^{\alpha'}-k_{n}^{\alpha'}-k_{s}^{\alpha'}\right)x_{\alpha'}}e^{i\left(k_{a}^{\alpha'}+k_{c}^{\alpha'}-k_{b}^{\alpha'}-k_{d}^{\alpha'}\right)x_{\alpha'}}d^{4}x\\
 & =-a_{b}^{\dagger}a_{d}^{\dagger}a_{a}a_{c}\left(k_{a}^{0'}-k_{b}^{0'}\right)\left(k_{a}^{j'}-k_{b}^{j'}\right),
\end{align}
Although this equation looks complicated, it can be understood quite
simply. The derivative terms $\partial_{0'}\partial_{j'}$ pull out
the factor $i^{2}\left(k_{n}^{0'}-k_{m}^{0'}\right)\left(k_{n}^{j'}-k_{m}^{j'}\right)$
from the Fourier decomposition of $\rho$. Then the integration over
$e^{i\left(k_{a}^{\alpha'}+k_{c}^{\alpha'}-k_{b}^{\alpha'}-k_{d}^{\alpha'}\right)x_{\alpha'}}$
selects the combination of wave-vectors $k_{a}^{\alpha'},k_{c}^{\alpha'},k_{b}^{\alpha'},k_{d}^{\alpha'}$
due to the orthonormality of the Fourier basis. Moving onto the second
term, a similar process shows:
\begin{align}
\int\partial_{0'}\rho\partial_{j'}\rho e^{i\left(k_{a}^{\alpha'}+k_{c}^{\alpha'}-k_{b}^{\alpha'}-k_{d}^{\alpha'}\right)x_{\alpha'}}d^{4}x & =-a_{b}^{\dagger}a_{d}^{\dagger}a_{a}a_{c}\left(k_{a}^{0'}-k_{b}^{0'}\right)\left(k_{c}^{j'}-k_{d}^{j'}\right),
\end{align}
Consequently, by {[}Eq. \ref{eq:4.23-1}{]}: 
\begin{align}
\int\rho^{2}\partial_{0'}\partial_{j'}\log\rho e^{i\left(k_{a}^{\alpha'}+k_{c}^{\alpha'}-k_{b}^{\alpha'}-k_{d}^{\alpha'}\right)x_{\alpha'}}dx= & a_{b}^{\dagger}a_{d}^{\dagger}a_{a}a_{c}\left(k_{a}^{0'}-k_{b}^{0'}\right)\left[\left(k_{a}^{j'}-k_{b}^{j'}\right)-\left(k_{c}^{j'}-k_{d}^{j'}\right)\right]
\end{align}
This quantity must be zero for $\rho$ to be log-separable. Therefore
if $\left(k_{a}^{0'}-k_{b}^{0'}\right)\neq0$ we must have instead:
\begin{equation}
\left(k_{a}^{j'}-k_{b}^{j'}\right)-\left(k_{c}^{j'}-k_{d}^{j'}\right)=0,
\end{equation}
since it is assumed we have chosen $a,b,c,d$ such that the factor
$a_{b}^{\dagger}a_{d}^{\dagger}a_{a}a_{c}\neq0.$ To simplify this
further, choose $b=d$, then the condition becomes: 
\begin{equation}
k_{a}^{j'}=k_{c}^{j'}.
\end{equation}
Repeating the procedure for $j'=1'$, $j'=2'$, $j'=3'$ gives $k_{a}^{1'}=k_{c}^{1'}$,
$k_{a}^{2'}=k_{c}^{2'}$, $k_{a}^{3'}=k_{c}^{3'}$. The on mass-shell
condition then specifies the last degree of freedom, and hence $k_{a}^{0'}=k_{c}^{0'}$.
Consequently, it must be the case that $k_{a}^{0'}=k_{c}^{0'}$ for
any $a\neq c$ with $a_{a},a_{c}\neq0$. To understand this result
further, the condition $k_{a}^{0'}=k_{c}^{0'}$ for all $a_{a},a_{c}\neq0$
is equivalent to $k_{n}^{0'}=E$ for all non-zero $a_{n}$, where
$E$ is some fixed constant which we identify as the energy. In other
words, the wavefunction is constructed from Fourier modes which share
the same energy state. The wavefunction is in an energy eigenstate,
which can be seen by taking the time-derivative of $\psi$:
\begin{align}
i\hbar\partial_{0'}\psi= & \sum_{n}-i^{2}k_{n}^{0'}a_{n}e^{-ik_{n}^{\alpha'}x_{\alpha'}}\\
= & E\sum_{n}a_{n}e^{-ik_{n}^{\alpha'}x_{\alpha'}}\\
= & E\psi.
\end{align}
Therefore the meaning of the inertial reference frame comes into view.
The inertial reference frame corresponds to a stationary probability
density $\partial_{0'}\rho=0$ and an energy eigenstate $\psi$. Furthermore,
$\frac{\partial_{0'}S}{\hbar}=E=\text{constant}$, which represents
the fact that the inertial reference frame defines a clock ticking
at a constant rate proportional to proper time. The constant of energy
$E$ will be related to the rest mass of the particle. 

The demonstration of the constant rate of $\partial_{0'}S$ over the
spatial coordinates is important, because it might be believed that
the relativistic quantum Hamilton-Jacobi equation in the inertial
reference frame implies $\partial_{0'}S$ is a function of the spatial
coordinates ${\bf x}$ due to the presence of the quantum potential
term, in variance to common intuition from classical relativistic
mechanics. However, it appears that $\frac{\partial_{0'}S}{\hbar}=E$
is indeed not a function of ${\bf x}'$, which is in agreement with
the classical result. 

In practice, the intuitive result of $\frac{\partial_{0'}S}{\hbar}=E$
is enabled by the special form of the quantum potential, in particular
that it is separable in the inertial reference frame since $\rho$
is log-separable, and additionally that it has no time-dependence
in the inertial frame as $\partial_{0'}\rho=0$. This will be discussed
further in section \ref{subsec:QHJ_in_inertial}. 

\subsection{\label{subsec:QHJ_in_inertial}Quantum Hamilton-Jacobi equation in
the stationary reference frame}

Another piece of evidence for the condition $\partial_{0'}\rho=0$
comes from the quantum Hamilton-Jacobi equation in the rest frame.
Because the quantum potential has the following alternative form {[}Eq.
\ref{eq:C.8}{]}: 
\begin{equation}
Q=-\frac{\hbar^{2}}{4m}\partial_{\mu}\partial^{\mu}\log\rho-\frac{\hbar^{2}}{8m}\partial_{\mu}\log\rho\partial^{\mu}\log\rho,
\end{equation}
the quantum potential is separable since $\log\rho$ is separable.
Using $\rho=e^{g(x_{0'})+g_{{\bf x'}}({\bf x}')}$ the quantum potential
can be written as:
\begin{equation}
Q=Q_{0'}(x_{0'})+Q_{{\bf x'}}({\bf x}'),
\end{equation}
where:
\begin{align}
Q_{0'}(x_{0'})= & -\frac{\hbar^{2}}{4m}\partial_{0'}\partial^{0'}g(x_{0'})-\frac{\hbar^{2}}{8m}\partial_{0'}g(x_{0'})\partial^{0'}g(x_{0'}),\label{eq:4.32}
\end{align}
\begin{align}
Q_{{\bf x'}}({\bf x}')= & -\frac{\hbar^{2}}{4m}\partial_{j'}\partial^{j'}g_{{\bf x}'}({\bf x}')-\frac{\hbar^{2}}{8m}\partial_{j'}g_{{\bf x}'}({\bf x}')\partial^{j'}g_{{\bf x}'}({\bf x}').
\end{align}
Turning to the quantum Hamilton-Jacobi equation, using the metric
signature $\eta=(1,-1,-1,-1)$, in the rest frame it becomes:
\begin{equation}
\frac{1}{2m}\partial_{0'}S\partial_{0'}S-V-Q_{0'}(x_{0'})-Q_{{\bf x'}}({\bf x}')-\frac{1}{2}mc^{2}=0.\label{eq:4.33}
\end{equation}
Taking the gradient with respect to $\partial_{j'}$ where $j'\in[1',2',3']$
gives: 
\begin{equation}
\partial_{j'}\left[V+Q_{{\bf x'}}({\bf x}')\right]=0.
\end{equation}
Therefore:
\begin{equation}
V+Q_{{\bf x'}}({\bf x}')=f(x_{0'}),
\end{equation}
which implies that $V$ is also separable:
\begin{equation}
V=V_{0'}(x_{0'})+V_{{\bf x'}}({\bf x}').
\end{equation}
Consequently, we know that: 
\begin{equation}
V_{{\bf x'}}({\bf x}')+Q_{{\bf x'}}({\bf x}')=\alpha,
\end{equation}
where $\alpha$ is a constant, because differentiating {[}Eq. \ref{eq:4.33}{]}
with respect to $\partial_{j'}$ gives:
\begin{align}
\partial_{j'}\left[V_{{\bf x'}}({\bf x}')+Q_{{\bf x'}}({\bf x}')\right]= & \partial_{j'}\alpha=0.
\end{align}
The quantum Hamilton-Jacobi equation then becomes:
\begin{equation}
\frac{1}{2m}\partial_{0'}S\partial_{0'}S-V_{0'}(x_{0'})-Q_{0'}(x_{0'})-\frac{1}{2}mc^{2}-\alpha=0.
\end{equation}
Since the continuity equation can be solved to give $\rho\partial_{0'}S=f({\bf x}')$
and $\partial_{0'}S>0$ we must also have: 
\begin{equation}
\partial_{0'}S=e^{-g_{0'}(x_{0'})+a({\bf x}')},
\end{equation}
where $a({\bf x}')=e^{\log f({\bf x}')}$. Substituting in $\partial_{0'}S=e^{-g(x_{0})+a}$
and $Q_{0'}(x_{0'})$ given by {[}Eq. \ref{eq:4.32}{]}:
\begin{equation}
\frac{1}{2m}e^{-2g(x_{0'})+2a}-V_{0'}(x_{0'})+\frac{\hbar^{2}}{4m}\partial_{0'}\partial^{0'}g(x_{0'})+\frac{\hbar^{2}}{8m}\partial_{0'}g(x_{0'})\partial^{0'}g(x_{0'})-\frac{1}{2}mc^{2}-\alpha=0.\label{eq:4.39}
\end{equation}
This is an autonomous differential equation for $g(x_{0'})$. Due
to the structure of autonomous differential equations, they have very
restrictive solutions. Such restrictions impose unphysical and non-local
constraints on $\rho$ which are difficult to satisfy unless $\partial_{0'}\log\rho=\partial_{0'}g(x_{0'}^{b})=0$.
However if $\partial_{0'}\rho=0$ as demonstrated in section \ref{subsec:Fourier-mode-decomposition},
this implies $\partial_{0'}\log\rho=0$. Therefore:
\begin{equation}
Q_{0'}(x_{0'})=0,
\end{equation}
resulting in the equation: 
\begin{equation}
\frac{1}{2m}e^{-2g(x_{0'})+2c}-V_{0'}(x_{0'})-\frac{1}{2}mc^{2}-\alpha=0,
\end{equation}
which is a differential equation that can easily be solved for $g(x_{0'})$.
Furthermore the equation also can be written as: 
\begin{equation}
\frac{1}{2m}\partial_{0'}S\partial^{0'}S-V_{0'}(x_{0'})-\frac{1}{2}mc^{2}+\alpha=0,
\end{equation}
therefore:
\begin{align}
\frac{\partial_{0'}S}{m}= & c\sqrt{1+\frac{2\left(V_{0'}+\alpha\right)}{mc^{2}}}.\label{eq:4.46}
\end{align}
Note that $\alpha$ is a constant not dependent on ${\bf x}'$, so
this quantity demonstrates that $\frac{\partial_{0'}S}{m}$ ticks
at a constant rate if the potential $V_{0}$ is a constant. Letting
$\alpha=0$ produces an even more interesting analogy as it explains
the existence of the $V_{0'}$ term in the formula:
\begin{equation}
\frac{\partial_{0'}S}{m}=c\sqrt{1+\frac{2V_{0'}}{mc^{2}}}.
\end{equation}
The quantity $\frac{\partial_{0'}S}{m}d\tau$ is the rate of ticking
of an internal clock in the rest frame. It is equal to $cd\tau$ multiplied
by a time-dilation factor due to the $g_{0'0'}=1+\frac{2V_{0'}}{m^{2}c^{2}}$
component of the spacetime metric, where $V_{0'}$ might be for example,
the gravitational potential energy. However, if $\alpha\neq0$, {[}Eq.
\ref{eq:4.46}{]} suggests that the spatial part of the quantum potential
may also contribute to the metric component $g_{0'0'}$. 

\section{\label{sec:Momentum-equivariance}Momentum equivariance}

\subsection{\label{subsec:Position-equivarance}Position equivariance }

In sections \ref{sec:Stochastic-mechanics} and \ref{sec:Cauchy-momentum-equations}
we derived two classical-like equations of motion {[}Eq. \ref{eq:3.23-1}{]}
and {[}Eq. \ref{eq:3.22}{]}:
\begin{equation}
\frac{D\rho}{D\tau}+\rho\left(\frac{1}{m}\partial_{\mu}\partial^{\mu}S'\right)=0,\label{eq:5.1}
\end{equation}

\begin{equation}
\frac{D\rho\partial_{\nu}S'}{D\tau}+\rho\partial_{\nu}S'\left(\frac{1}{m}\partial_{\mu}\partial^{\mu}S'\right)+\partial_{\nu}V=0,
\end{equation}
which describe how the probability density and momentum update along
the trajectory of stochastic particles. Suppose there exists a de
Broglie-Bohm particle corresponding to this system of equations. If
we represent the particle by the delta function $\delta(x^{\mu}-X_{\tau}^{\mu})$,
it is known that this function is constant in the reference frame
of the particle. The stochastic Lagrangian derivative of the delta
function is zero as the Lagrangian derivative describes the propagation
of the quantity in the co-moving reference frame. Therefore:
\begin{equation}
\frac{D\delta(x-X_{\tau}^{a})}{D\tau}=0.
\end{equation}
Add to this equation an additional term:
\begin{equation}
\delta(x-X_{\tau}^{a})\partial_{\mu}\frac{DX^{a\mu}}{D\tau}=0,
\end{equation}
which is zero because $\frac{DX^{\mu}(\tau)}{D\tau}$ is only explicitly
a function of $\tau$ (or alternatively use the phase space approach
discussed in\emph{ }Appendix \ref{subsec:Equivariance-in-the}). Therefore:
\begin{equation}
\frac{D\delta(x-X_{\tau}^{a})}{D\tau}+\delta(x-X_{\tau}^{a})\partial_{\mu}\frac{DX^{a\mu}}{D\tau}=0.
\end{equation}
By writing out the stochastic Lagrangian derivative $\frac{D}{D\tau}$
explicitly:
\begin{align}
0= & \frac{\partial\delta(x-X_{\tau}^{a})}{\partial\tau}+\frac{DX^{a\mu}}{D\tau}\partial_{\mu}\delta(x-X_{\tau}^{a})-\frac{\hbar}{2m}\partial_{\mu}\partial^{\mu}\left(\delta(x-X_{\tau}^{a})\right)+\delta(x-X_{\tau}^{a})\partial_{\mu}\frac{DX^{a\mu}}{D\tau}\\
= & \frac{\partial\delta(x-X_{\tau}^{a})}{\partial\tau}+\partial_{\mu}\left(\delta(x-X_{\tau}^{a})\frac{DX^{a\mu}}{D\tau}\right)-\frac{\hbar}{2m}\partial_{\mu}\partial^{\mu}\left(\delta(x-X_{\tau}^{a})\right).
\end{align}
Averaging over the particles of the statistical ensemble then gives:
\begin{equation}
\frac{\partial\underset{N\rightarrow\infty}{\lim}\frac{1}{N}\sum_{a=1}^{N}\delta(x-X_{\tau}^{a})}{\partial\tau}+\partial_{\mu}\left(\underset{N\rightarrow\infty}{\lim}\frac{1}{N}\sum_{a=1}^{N}\delta(x-X_{\tau}^{a})\frac{\partial^{\mu}S'}{m}\right)-\frac{\hbar}{2m}\partial_{\mu}\partial^{\mu}\left(\underset{N\rightarrow\infty}{\lim}\frac{1}{N}\sum_{a=1}^{N}\delta(x-X_{\tau}^{a})\right)=0,
\end{equation}
which equals: 
\begin{equation}
\frac{\partial\rho(x,\tau)}{\partial\tau}+\partial_{\mu}\left(\rho(x,\tau)\frac{\partial^{\mu}S'}{m}\right)-\frac{\hbar}{2m}\partial_{\mu}\partial^{\mu}\rho(x,\tau)=0.\label{eq:5.9}
\end{equation}
In this derivation we have used the condition: 
\begin{equation}
\frac{DX^{a\mu}}{D\tau}=\frac{\partial^{\mu}S'}{m},
\end{equation}
which is just the guidance equation along the particle trajectory.
We have also identified the particle density to be: 
\begin{equation}
\rho(x,\tau)=\underset{N\rightarrow\infty}{\lim}\frac{1}{N}\sum_{a=1}^{N}\delta(x-X_{\tau}^{a}),
\end{equation}
where $N$ is the number of particle configurations in the statistical
ensemble. {[}Eq. \ref{eq:5.9}{]} is the same as the transformed continuity
equation {[}Eq. \ref{eq:2.18}{]}, therefore the statistical ensemble
of de Broglie-Bohm particles satisfies the equivariance property for
their positions. 

\subsection{\label{subsec:Momentum-equivarance-1}Momentum equivariance \label{subsec:Momentum-equivarance}}

We now make the ansatz that the equation for the acceleration along
particle trajectories is:
\begin{equation}
\frac{D}{D\tau}\left(\frac{DX_{\tau}^{a}}{D\tau}\right)=\frac{D\frac{1}{m}\partial_{\nu}S'|_{x=X_{\tau}^{a}}}{D\tau}=-\frac{1}{m}\partial_{\nu}V(x)|_{x=X_{\tau}^{a}}\label{eq:5.12}
\end{equation}
which is equivalent to Newton's law in the Lagrangian reference frame
of the particle configuration. The function $\partial_{\nu}S'|_{x=X_{\tau}^{a}}$
and $\partial_{\nu}V(x)|_{x=X_{\tau}^{a}}$ can alternatively be written
using delta functions as:
\begin{equation}
\partial_{\nu}S'|_{x=X_{\tau}^{a}}=\delta(x-X_{\tau}^{a})\partial_{\nu}S'(x),
\end{equation}

\begin{equation}
\partial_{\nu}V(x)|_{x=X_{\tau}^{a}}=\delta(x-X_{\tau}^{a})\partial_{\nu}V.
\end{equation}
Therefore {[}Eq. \ref{eq:5.12}{]} is equivalent to:
\begin{equation}
\frac{D\delta(x-X_{\tau}^{a})\partial_{\nu}S'}{D\tau}=-\delta(x-X_{\tau}^{a})\partial_{\nu}V.
\end{equation}
An additional term $\delta(x-X_{\tau}^{a\mu})\partial_{\nu}S'\partial_{\mu}\left(\frac{DX^{a\mu}}{D\tau}\right)=0$
can be added to this equation. This additional term is zero because
$\frac{DX^{a}}{D\tau}$ is explicitly a function of time only (or
alternatively use the phase space approach discussed in Appendix \ref{subsec:Equivariance-in-the}).
Therefore:
\begin{equation}
\frac{D\delta(x-X_{\tau}^{a})\partial_{\nu}S'}{D\tau}+\delta(x-X_{\tau}^{a})\partial_{\nu}S'\partial_{\mu}\left(\frac{DX^{a\mu}}{D\tau}\right)+\delta(x-X_{\tau}^{a})\partial_{\nu}V=0,
\end{equation}
or equivalently by writing $\frac{D}{D\tau}$ explicitly:
\begin{align}
0= & \frac{\partial\delta(x-X_{\tau}^{a})\partial_{\nu}S'}{\partial\tau}+\frac{DX_{\tau}^{a\mu}}{D\tau}\partial_{\mu}\left(\delta(x-X_{\tau}^{a})\partial_{\nu}S'\right)+\delta(x-X_{\tau}^{a})\partial_{\nu}S'\partial_{\mu}\left(\frac{DX^{a\mu}}{D\tau}\right)\ldots\nonumber \\
 & \ldots-\frac{\hbar}{2m}\partial_{\mu}\partial^{\mu}\left(\delta(x-X_{\tau}^{a})\partial_{\nu}S'\right)+\delta(x-X_{\tau}^{a})\partial_{\nu}V\\
= & \frac{\partial\delta(x-X_{\tau}^{a})\partial_{\nu}S'}{\partial\tau}+\partial_{\mu}\left(\delta(x-X_{\tau}^{a})\partial_{\nu}S'\frac{DX_{\tau}^{a\mu}}{D\tau}\right)-\frac{\hbar}{2m}\partial_{\mu}\partial^{\mu}\left(\delta(x-X_{\tau}^{a})\partial_{\nu}S'\right)+\delta(x-X_{\tau}^{a})\partial_{\nu}V.
\end{align}
Averaging over particles of the statistical ensemble gives:
\begin{align}
0= & \frac{\partial\underset{N\rightarrow\infty}{\lim}\frac{1}{N}\sum_{a=1}^{N}\delta(x-X_{\tau}^{a})\partial_{\nu}S'}{\partial\tau}+\partial_{\mu}\left(\underset{N\rightarrow\infty}{\lim}\frac{1}{N}\sum_{a=1}^{N}\delta(x-X_{\tau}^{a})\partial_{\nu}S'\frac{DX_{\tau}^{a\mu}}{D\tau}\right)\ldots\nonumber \\
 & \ldots-\frac{\hbar}{2m}\partial_{\mu}\partial^{\mu}\left(\underset{N\rightarrow\infty}{\lim}\frac{1}{N}\sum_{a=1}^{N}\delta(x-X_{\tau}^{a})\partial_{\nu}S'\right)+\underset{N\rightarrow\infty}{\lim}\frac{1}{N}\sum_{a=1}^{N}\delta(x-X_{\tau}^{a})\partial_{\nu}V,
\end{align}
therefore:
\begin{equation}
\frac{\partial\rho\partial_{\nu}S'}{\partial\tau}+\partial_{\mu}\left(\rho\partial_{\nu}S'\frac{DX^{\mu}}{D\tau}\right)+\rho\partial_{\nu}V=\frac{\hbar}{2m}\partial_{\mu}\partial^{\mu}\left(\rho\partial_{\nu}S'\right).\label{eq:5.16}
\end{equation}
Consequently setting $\frac{DX^{a\mu}}{D\tau}=\frac{\partial^{\mu}S'}{m}$
(which is the de Broglie-Bohm guidance equation) implies by {[}Eq.
\ref{eq:5.16}{]} that the particle momentum satisfies the same Cauchy-momentum
equation as the quantum mechanical momentum current density {[}Eq.
\ref{eq:3.19}{]}. Therefore, the de Broglie-Bohm particles satisfy
an equivariance relationship for the momenta. 

The equivariance relationship proves that the ansatz {[}Eq. \ref{eq:5.12}{]}
is correct. If the de Broglie-Bohm particles update their velocity
according to {[}Eq. \ref{eq:5.12}{]}, then evidently they will reproduce
the quantum predictions because i) the statistical ensemble of particles
will produce the correct momenta $\rho\partial_{\nu}S'$ via this
momentum equivariance relationship ii) the statistical ensemble will
produce the correct probability density $\rho$ through the position
equivariance relationship shown in section \ref{subsec:Position-equivarance}
and iii) since $\rho\partial_{\nu}S'$ and $\rho$ are correct, then
the velocity $\partial_{\nu}S'=\frac{\rho\partial_{\nu}S'}{\rho}$
will be correct. 

\subsection{Path independence}

In sections \ref{subsec:Position-equivarance} and \ref{subsec:Momentum-equivarance-1}
, we have assumed that:
\begin{equation}
\underset{N\rightarrow\infty}{\lim}\frac{1}{N}\sum_{a=1}^{N}\delta(x-X_{t}^{a})\frac{DX^{a\mu}}{D\tau}=\rho\frac{DX^{\mu}}{D\tau},\label{eq:7.27}
\end{equation}
and: 
\begin{equation}
\underset{N\rightarrow\infty}{\lim}\frac{1}{N}\sum_{a=1}^{N}\delta(x-X_{t}^{a})\partial_{\nu}S'(x)\frac{DX^{a\mu}}{D\tau}=\rho\partial_{\nu}S'(x)\frac{DX^{\mu}}{D\tau}.
\end{equation}
These equations are worth analysing in more detail. The first equation
states that the normalised sum over velocities of the statistical
ensemble gives the average velocity $\rho\frac{DX^{\mu}}{D\tau}$.
The second equation follows from the first because: 
\begin{align}
\underset{N\rightarrow\infty}{\lim}\frac{1}{N}\sum_{a=1}^{N}\delta(x-X_{t}^{a})\partial_{\nu}S'(x)\frac{DX^{a\mu}}{D\tau}= & \partial_{\nu}S'(x)\underset{N\rightarrow\infty}{\lim}\frac{1}{N}\sum_{a=1}^{N}\delta(x-X_{t}^{a})\frac{DX^{a\mu}}{D\tau},
\end{align}
as $\partial_{\nu}S'(x)$ does not depend on the particle label $a$.
It is important to note that since $\partial_{\nu}S'(x)$ is updated
along the particle trajectory according to a classical-like equation
of motion involving a conservative force $\partial_{\nu}V$, its calculated
value is path independent.

Conservative forces produce path independence at the level of Newtonian-like
equations of motion for velocity and momentum. This is equally true
for classical Newtonian mechanics as it is for deterministic Bohmian
mechanics. It is also true for classical stochastic dynamics equations
of motion, and the stochastic version of Bohmian mechanics developed
in this paper. 

A key thing to recognize about the stochastic guidance equations is
that only the particle position is directly effected by the random
motion. The second-order Newtonian law for updating the velocity is
only affected indirectly insofar as the particles take a different
path, however the mathematical equation for the second order dynamics
contains no stochastic terms directly. 

The fortuitous consequence of the path independence is that all particles
of the statistical ensemble will have the same value of $\partial_{\nu}S'(x)$
when they are located at the same spacetime coordinates, assuming
these particles originated with the same value for $\partial_{\nu}S'(x)$.
This is to say that we can assume $\frac{DX^{a\mu}}{D\tau}$ all equal
a single value $\frac{DX^{\mu}}{D\tau}=\partial^{\mu}S'(x)$ when
evaluated at the same position $X_{t}^{a}=x$, which gives {[}Eq.
\ref{eq:7.27}{]}. 

Now a counterargument could be made that if the force $\partial_{\nu}V$
is time dependent, this would break the path independence. For example,
an ensemble of particles could be separated into two groups. For one
group a conservative force could be applied and then removed before
bringing the two groups of particles back together. This allows the
particles to accumulate a difference in momentum. In this thought
experiment, the time-dependent nature of the force disrupts the path
independence. However, we must recognize that in the relativistic
domain, the potential $V$ is a static quantity in the background
of proper time. Any time dependence is in the local-time parameter
$t$ not the proper time $\tau$. Therefore the potential $V$ is
time-independent from the perspective of proper time $\tau$, and
so the force $\partial_{\nu}V$ is conservative from the perspective
of dynamics in $\tau$. 

\section{Conservation of momentum }

\subsection{Deterministic mechanics }

From the quantum mechanical equations of motion for $\psi$, the continuity
equation, Hamilton-Jacobi equation and Cauchy-momentum equation can
be derived. These represent conservation of probability density, energy
density and momentum density respectively. Only two of these three
equations are actually required however, as the third can be derived
from the first two. The de Broglie-Bohm interpretation is typically
formulated in terms of the continuity and Hamilton-Jacobi equations,
while the Cauchy-momentum equation is ignored in the interpretation.
Ordinarily this is unproblematic, because for deterministic dynamics,
the solutions to the three equation are consistent along the trajectories
of the particles. For instance, in the non-relativistic case of Bohmian
mechanics the continuity equation is: 
\begin{equation}
\frac{d\rho}{dt}=-\rho\nabla^{2}S,
\end{equation}
while the gradient of the quantum Hamilton-Jacobi equation is:
\begin{equation}
\frac{d\nabla S}{dt}=-\nabla V-\nabla Q,
\end{equation}
and the quantum Cauchy-momentum equation is:
\begin{equation}
\frac{d\rho\nabla S}{dt}=-\left(\rho\nabla S\right)\nabla^{2}S-\rho\nabla V-\rho\nabla Q.
\end{equation}
Evidently, from these equations we have the relationship:
\begin{equation}
\frac{d\rho\nabla S}{dt}=\rho\frac{d\nabla S}{dt}+\frac{d\rho}{dt}\nabla S,\label{eq:7.4}
\end{equation}
which implies that up to order $\delta t^{2}$:
\begin{equation}
\left[\rho\nabla S\right]_{t-1}+\delta t\frac{d\rho\nabla S}{dt}=\left(\left[\rho\right]_{t-1}+\delta t\frac{d\rho}{dt}\right)\left(\left[\nabla S\right]_{t-1}+\delta t\frac{d\nabla S}{dt}\right),\label{eq:6.7}
\end{equation}
or equivalently dividing by mass $m$:
\begin{equation}
\left[\rho\frac{\nabla S}{m}\right]_{t}=\left[\rho\right]_{t}\left[\frac{\nabla S}{m}\right]_{t},\label{eq:6.7-1}
\end{equation}
which demonstrates the consistency of updating the momentum $\rho\frac{\nabla S}{m}$
with separately updating the value of the probability density $\rho$
and velocity $\frac{\nabla S}{m}$.

\subsection{Stochastic mechanics}

This trivial result for the consistency of deterministic particle
dynamics breaks down in the stochastic case, due to the non-linear
term of the Itô calculus present in the stochastic Lagrangian derivative.
In the relativistic stochastic case, the three dynamical equations
are the continuity equation {[}Eq. \ref{eq:3.23-1}{]}:
\begin{equation}
\frac{D\rho}{D\tau}=-\rho\left(\frac{1}{m}\partial_{\mu}\partial^{\mu}S'\right),\label{eq:6.7-2}
\end{equation}
the gradient of the quantum Hamilton-Jacobi equation given by (Appendix
\ref{sec:Quantum-Hamilton-Jacobi-equation}\emph{ }{[}Eq. \ref{eq:e16}{]}):
\begin{align}
\frac{D\partial_{\nu}S'}{D\tau}= & -\partial_{\nu}V-2\partial_{\nu}Q-\frac{\hbar}{m}\partial_{\mu}\partial^{\mu}\partial_{\nu}S',\label{eq:6.8}
\end{align}
and the quantum Cauchy-momentum equation {[}Eq. \ref{eq:3.22}{]}:
\begin{equation}
\frac{D\rho\partial_{\nu}S'}{D\tau}=-\rho\partial_{\nu}S'\left(\frac{1}{m}\partial_{\mu}\partial^{\mu}S'\right)-\rho\partial_{\nu}V.\label{eq:6.9}
\end{equation}
Evidently, the consistency condition is no longer satisfied, i.e.:
\begin{equation}
\frac{D\rho\partial_{\nu}S'}{D\tau}\neq\rho\frac{D\partial_{\nu}S'}{D\tau}+\frac{D\rho}{D\tau}\partial_{\nu}S'.\label{eq:6.14}
\end{equation}
There is an additional term present $-2\partial_{\nu}Q-\frac{\hbar}{m}\partial_{\mu}\partial^{\mu}\partial_{\nu}S'$,
which can be seen by substituting {[}Eq. \ref{eq:6.7-2}, \ref{eq:6.8},
\ref{eq:6.9}{]} into {[}Eq. \ref{eq:6.14}{]}. Therefore if $\rho$
is updated according to the continuity equation:
\begin{align}
\left[\rho\right]_{\tau}= & \left[\rho\right]_{\tau-1}+\delta\tau\frac{D\rho}{D\tau},
\end{align}
and $\partial_{\nu}S'$ is updated according to the gradient of the
quantum Hamilton-Jacobi equation:
\begin{equation}
\left[\partial_{\nu}S'\right]_{\tau}=\left[\partial_{\nu}S'\right]_{\tau-1}+\delta\tau\frac{D\partial_{\nu}S'}{Dt}.
\end{equation}
This is not the same as updating $\rho\partial_{\nu}S'$ according
to the quantum Cauchy-momentum equation:
\begin{align}
\left[\rho\partial_{\nu}S'\right]_{\tau}= & \left[\rho\partial_{\nu}S'\right]_{\tau-1}+\delta\tau\frac{D\rho\partial_{\nu}S'}{D\tau}.
\end{align}
In particular:
\begin{align}
\left[\rho\partial_{\nu}S'\right]_{\tau-1}+\delta\tau\frac{D\rho\partial_{\nu}S'}{D\tau}\neq & \left(\left[\rho\right]_{\tau-1}+\delta\tau\frac{D\rho}{D\tau}\right)\left(\left[\nabla S\right]_{\tau-1}+\delta\tau\frac{D\partial_{\nu}S'}{D\tau}\right),
\end{align}
which requires {[}Eq. \ref{eq:6.14}{]} to be satisfied. 

The implication is that for stochastic particles, an explicit choice
needs to be made as to whether the velocity $\frac{\partial_{\nu}S'}{m}$
is calculated from the quantum Hamilton-Jacobi equation, or whether
the momentum $\rho\partial_{\nu}S'$ is calculated from the Cauchy-momentum
equations. The choice the traditional de Broglie-Bohm interpretation
makes of using the quantum Hamilton-Jacobi equation may not be valid
in the stochastic version of the theory. While a form of Newton's
second law (which is derived from the gradient of the quantum Hamilton-Jacobi
equation) is upheld for deterministic Bohmian mechanics, it breaks
down in the stochastic case where it needs to be replaced with the
concept of momentum equivariance discussed in section \ref{sec:Momentum-equivariance}. 

Since equivariance for particle positions is the crucial element of
the de Broglie-Bohm interpretation that allows the continuity equation
to hold true, we might suppose that momentum equivariance is of similar
importance as it ensures the Cauchy-momentum equations hold true.
These two equivariance relationships point toward placing special
emphasis on position and momentum conservation, not on the particle
velocity, which may not be fundamental. 

Examining {[}Eq. \ref{eq:6.8}{]} shows that in addition to the force
term of the quantum potential $-2\partial_{\nu}Q$ being doubled in
the equation for acceleration, there is an extra force term $-\partial_{\nu}\left(\frac{\hbar}{m}\partial_{\mu}\partial^{\mu}S'\right)$.
Meanwhile, the quantum Cauchy-momentum equations {[}Eq. \ref{eq:6.9}{]}
transform into a classical-like representation where the quantum potential
is absent. This suggests that the quantum Cauchy-momentum equations
are the natural choice. The stochastic transformation appears to have
shunted all of the problems of the quantum Cauchy-momentum equations,
i.e. the non-local quantum potential, into quantum Hamilton-Jacobi
equations. This is why the quantum potential term is doubled in the
transformed quantum Hamilton-Jacobi equations, but absent in the transformed
quantum Cauchy-momentum equations. 

\subsection{Consistency condition\label{subsec:Consistency-condition}}

As discussed above, all three quantities of probability density, velocity
and momentum density cannot be simultaneously updated in the reference
frame of stochastic particles using the standard equations of motion.
This is true unless a hidden condition $\partial_{\mu}\rho\partial_{\nu}\partial^{\mu}S'=0$
is satisfied. This hypothetical condition arises because, by the definition
of the stochastic Lagrangian derivative, it is a mathematical identity
that: {\small{}
\begin{align}
\frac{D\rho\partial_{\nu}S'}{D\tau}= & \frac{\partial\rho\partial_{\nu}S'}{\partial t}+\frac{1}{m}\partial_{\mu}S'\partial_{\mu}\left(\rho\partial_{\nu}S'\right)-\frac{\hbar}{2m}\partial_{\mu}\partial^{\mu}\left(\rho\partial_{\nu}S'\right)\label{eq: 6.16}\\
= & \left(\frac{\partial\rho}{\partial t}+\frac{1}{m}\partial_{\mu}S'\partial_{\mu}\rho-\frac{\hbar}{2m}\partial_{\mu}\partial^{\mu}\rho\right)\partial_{\nu}S'+\rho\left(\frac{\partial\partial_{\nu}S'}{\partial t}+\frac{1}{m}\partial_{\mu}S'\partial_{\mu}\partial_{\nu}S-\frac{\hbar}{2m}\partial_{\mu}\partial^{\mu}\partial_{\nu}S'\right)-\frac{\hbar}{m}\partial_{\mu}\rho\partial^{\mu}\partial_{\nu}S'\\
= & \rho\frac{D\partial_{\nu}S'}{D\tau}+\frac{D\rho}{D\tau}\partial_{\nu}S'-\frac{\hbar}{m}\partial_{\mu}\rho\partial^{\mu}\partial_{\nu}S',\label{eq:7.12}
\end{align}
}where the additional term $-\frac{\hbar}{m}\partial_{\mu}\rho\partial^{\mu}\partial_{\nu}S'$
is equivalent to the non-linear component of the product rule for
Itô calculus. It can be shown that this equation holds true by substituting
the the equations for $\frac{D\rho}{D\tau}$, $\frac{D\partial_{\nu}S'}{D\tau}$
and $\frac{D\rho\partial_{\nu}S'}{D\tau}$ given by {[}Eq. \ref{eq:6.7-2},
\ref{eq:6.8}, \ref{eq:6.9}{]}, and using the conservation condition
$\partial_{\mu}\left(\rho\partial^{\mu}\partial_{\nu}S\right)=0$.
In particular, it can be shown after some work that: 

\begin{equation}
-2\partial_{\nu}Q-\frac{\hbar}{m}\partial_{\mu}\partial^{\mu}\partial_{\nu}S=\frac{\hbar}{m}\partial_{\mu}\rho\partial^{\mu}\partial_{\nu}S',\label{eq:6.18}
\end{equation}
which allows {[}Eq. \ref{eq:7.12}{]} to hold true. The derivation
of this equation is completed in Appendix \ref{sec:Consistency-condition}.
The successful derivation of {[}Eq. \ref{eq:6.18}{]} verifies the
correctness of the independent derivations for $\frac{D\rho\partial_{\nu}S'}{D\tau}$,
$\frac{D\partial_{\nu}S'}{D\tau}$ and $\frac{D\rho}{D\tau}$ in this
paper. It also shows how the Cauchy-momentum equation fits together
with the quantum Hamilton-Jacobi equation in {[}Eq. \ref{eq:7.12}{]}.
The equation for $\frac{D\partial_{\nu}S'}{D\tau}$ has two additional
terms $-2\rho\partial_{\nu}Q-\frac{\hbar}{m}\rho\partial_{\mu}\partial^{\mu}\partial_{\nu}S'$.
These turn out to be equal to $\frac{\hbar}{m}\partial_{\mu}\rho\partial^{\mu}\partial_{\nu}S'$
by {[}Eq. \ref{eq:6.18}{]}, which needs to be canceled by the $-\frac{\hbar}{m}\partial_{\mu}\rho\partial^{\mu}\partial_{\nu}S'$
term in {[}Eq. \ref{eq:7.12}{]} to make the equation balance. 

By equation {[}Eq. \ref{eq:7.12}{]}, the hypothetical condition $\partial_{\mu}\rho\partial_{\nu}\partial^{\mu}S'=0$
would force the stochastic Lagrangian derivative to act linearly $\frac{D\rho\partial_{\nu}S'}{D\tau}=\rho\frac{D\partial_{\nu}S'}{D\tau}+\frac{D\rho}{D\tau}\partial_{\nu}S'$
and therefore produce the same consistency as occurs in the deterministic
case. In our investigations, we searched for the existence of this
condition but were unable to prove the result. Note that in the rest
frame, if $\partial_{0'}\rho=0$ when $\partial S=(\partial^{0'}S,0,0,0)$
then $\partial_{0'}\rho\partial_{\nu'}\partial^{0'}S=0$. In this
case, the consistency condition is reduced to:
\begin{align}
\partial_{\mu'}\rho\partial_{\nu'}\partial^{\mu'}S'= & \frac{\hbar}{m}\partial_{0'}\rho\partial_{\nu'}\partial^{0'}S+\frac{\hbar^{2}}{2m}\partial_{\mu'}\rho\partial_{\nu'}\partial^{\mu'}\log\rho\\
= & \frac{\hbar^{2}}{2m}\partial_{\mu'}\rho\partial_{\nu'}\partial^{\mu'}\log\rho\\
= & \frac{\hbar^{2}}{4m}\rho\partial_{\nu'}\left(\partial_{\mu'}\log\rho\partial^{\mu'}\log\rho\right),
\end{align}
and so we see that the consistency condition is related to part of
the quantum potential, which has the form {[}Eq. \ref{eq:C.8}{]}:
\begin{equation}
\rho\partial_{\nu'}Q=-\frac{\hbar^{2}}{8m}\rho\partial_{\nu'}\left(\partial_{\mu'}\log\rho\partial^{\mu'}\log\rho\right)-\frac{\hbar^{2}}{4m}\rho\partial_{\mu}\partial^{\mu}\partial_{\nu'}\log\rho.
\end{equation}

\section{Philosophical discussion}

\subsection{Block universe is not static}

A common conceptualisation of the Block universe is that the matter
distribution is static and timeless in a physically real four-dimensional
spacetime. This is motivated for example in canonical general relativity
by the Hamiltonian constraint, for instance the Wheeler-DeWitt equation
$\hat{H}\Psi=0$ implies a static system with $\frac{\partial\Psi}{\partial\tau}=0$.
For static four-dimensional systems, traditional dynamics takes on
a new interpretation as a structural relationship between matter across
spacetime coordinates. 

The Klein-Gordon equation satisfies a Hamiltonian constraint, as evidenced
by the relativistic quantum Hamilton-Jacobi equation derived from
these equations, which is of the form of $\partial_{\tau}S+H=0$ with
$\partial_{\tau}S=0$ and $H=0$. However, the stochastic transformed
de Broglie-Bohm interpretation provides a new understanding in this
context, because contrary to the idea that matter under a Hamiltonian
constraint is static, there is continual motion of de Broglie-Bohm
particles generated by the stochastic guidance equations. This motion
occurs even in the deterministic case if $\partial_{\mu}S\neq0,$
but especially in the case of stochastic de Broglie-Bohm particles
due to the stochastic term present in the guidance equations. 

\subsection{\label{subsec:Dynamical-relaxation-to}Dynamical relaxation to quantum
equilibrium }

There is another way in which the de Broglie-Bohm interpretation can
have non-static dynamics in a Hamiltonian constrained system. As was
shown by Valentini \cite{key-3}, the de Broglie-Bohm interpretation
does not necessarily postulate the Born rule, because the Born rule
arises as a natural consequence of particle dynamics. This is the
concept of dynamical relaxation to quantum equilibrium, which is an
important and novel feature of the de Broglie-Bohm interpretation.
Any particle distribution out of equilibrium must get closer to the
quantum equilibrium distribution as the dynamics progresses. This
is an advantage to the de Broglie-Bohm interpretation as the Born-rule
can emerge from the dynamics instead of being postulated as a law
of quantum mechanics as an additional assumption. 

The concept of dynamical relaxation to quantum equilibrium can be
used to draw interesting conclusions. Suppose that at the initial
proper time $\tau_{0}$, the initial universe started out with a particle
distribution entirely localized on the local-time coordinate $t=0$.
In this example, the initial spatial coordinates can either specified
over range of the form $(0,x_{1},x_{2},x_{3})$ or alternatively localized
at a singularity with coordinates $(0,0,0,0)$. Due to dynamical relaxation
to quantum equilibrium, a distribution highly localized in the local-time
coordinate $t$ will spread out over a range of local-time coordinates
as the proper time progresses. 

Through this mechanism, we can resolve a point of concern regarding
determinism in Hamiltonian constrained system. The issue is that a
static Block universe implies a highly fine-tuned system, as there
is no motion and the final state is in some sense predetermined. However,
dynamical relaxation to quantum equilibrium ensures that this concern
is unwarranted. The initial conditions are not precisely specified,
but instead the quantum equilibrium distribution is reached on its
own accord through the process of dynamical relaxation. The final
state is not predetermined by the initial conditions, even though
the dynamics is inherently retrocausal. 

\subsection{Compactified local-time}

We furthermore propose a simple solution to the problem of how to
conceptualize four-dimensional spacetime in a realist manner. We argue
that the local-time coordinate $t$ is actually an internal degree
of freedom located at the spatial coordinates. In this interpretation
of the time coordinate, spacetime is not four-dimensional, it is three
dimensional with multiple times located at each coordinate of space. 

If there are no conceptual difficulties with understanding the phase
of a wavefunction as a `multivaluedness', it is conceivable that the
local-time coordinate can be understood in a similar manner. Retrocausality
can make sense from a local-realist perspective, because the time
coordinate becomes analogous to any other localized internal degree
of freedom such as the particle phase. Another point is that this
local-time degree of freedom does not have to be infinite, it is simple
to imagine that it is a closed loop which is periodic. This would
permit a cyclic universe. By contrast, it is difficult to conceptualise
time as having a closed topology and periodic nature if it is an extensive
variable similar to the spatial coordinates. To ensure local-causality,
it is sufficient to require that particles in this three-dimensional
space can only communicate if they i) have the same spatial coordinate
and ii) have the same value for the local-time degree of freedom. 

Given this perspective, retrocausal approaches may be a viable approach
toward Bell's theorem \cite{key-6} and entanglement correlations.
The main problem of retrocausality is not that the four-dimensional
spacetime coordinates cannot be understood in a realist ontology.
As argued above, this is simple to do by conceiving as time an internal
degree of freedom or multivaluedness. The real problems of retrocausality
are firstly the implicit fine-tuning of the initial conditions or
super-determinism, and secondly how to describe the system in three-dimensional
space instead of configuration space. 

However, in answer to the first problem of fine tuning, the stochastic
de Broglie-Bohm interpretation is not fine-tuned, as the quantum equilibrium
distribution can emerge naturally through the process of dynamical
relaxation. Furthermore, because the interpretation is stochastic
not deterministic, it does not have precise specification of the initial
conditions. In answer to the second problem of configuration space,
the stochastic de Broglie-Bohm interpretation developed within this
paper is local and can be interpreted in three-dimensional space,
due to the Newtonian form of the equations of motion for velocity
in the Lagrangian frame, which depend only upon the classical potential.

Therefore, the approach taken in this paper of having a local retrocausal
interpretation appears viable on these two measures. Furthermore,
it should be noted that the physical reality of spacetime coordinates
is already implicit within relativity theory, therefore it is not
a large logical step to presume this is the cause of quantum entanglement
also. Occam's razor applies to the situation in presuming these two
phenomena, of relativity and entanglement, are caused by the same
underlying mechanism of four-dimensional spacetime. 

\section{Conclusions}

\noindent Several new conceptions have been put forward in this paper.
Firstly, we have suggested that the quantum potential term is not
fundamental, as it can be removed from the relativistic quantum Cauchy-momentum
equations using the transform of the phase $S'=S+\frac{\hbar}{2}\log\rho$.
The resulting transformed Cauchy-momentum equations describe a Newtonian-like
law for updating particle momentum in the Lagrangian reference frame
of stochastic particles. This indicates a local-realist ontology can
be constructed for the Klein-Gordon system of equations.

Secondly, we have investigated the conservation of the rank-2 probability
current $\rho\partial_{\mu}\partial_{\nu}S$, which is seen to be
at the center of several issues. The non-conservation of this current
is related to non-locality in the non-relativistic Schrödinger system,
and conversely, conservation of this current in the Klein-Gordon system
enables the quantum potential term to be removed from the relativistic
quantum Cauchy-momentum equations under the stochastic transform of
the phase.

Thirdly, we have explored the consistency between the three equations
of motion for the quantum system; the quantum Hamilton-Jacobi equation,
continuity equation and quantum Cauchy-momentum equation. We have
shown that these equations are not mutually consistent in the Lagrangian
reference frame of stochastic particles, due to non-linear terms of
the Itô calculus occurring in the stochastic Lagrangian derivative.
Given that the quantum Cauchy-momentum equations become classical
under the stochastic transformation, we have suggested that these
equations should be regarded as fundamental instead of the quantum
Hamilton-Jacobi equations. 

In this respect, we have suggested that the approach of Bohmian and
Newtonian mechanics of describing the equations of motion in terms
of particle acceleration needs to be replaced by the concept of momentum
equivariance. Momentum equivariance enables the Cauchy-momentum equations
to determine how particle velocities are updated, instead of the gradient
of the Hamilton-Jacobi equation as in the case of Bohmian mechanics.
This may be an important clarification to how the second-order equations
of motion should be derived in complete generality to include the
case of stochastic particles. The picture of quantum mechanics we
have proposed has conservation of probability and conservation of
momentum playing fundamental roles, with position equivariance and
momentum equivariance as the core principles driving the interpretation. 

Finally, we have investigated the spacetime aspects of the interpretation.
We have suggested a simple way to understand spacetime coordinates,
with the local-time coordinate viewed as an internal degree of freedom
or multivaluedness available at each position in space. Furthermore,
we have proposed that dynamical relaxation to quantum equilibrium
explains how the stochastic de Broglie-Bohm interpretation can avoid
fine-tuning and superdeterminism despite being retrocausal in nature.
These both indicate that stochastic dynamics in four-dimensional spacetime
can provide an adequate resolution to the problem of non-locality.

\section*{Statements \& declarations}

\subsection*{Statement of originality}

All work is original research and is the sole research contribution
of the author. 

\subsection*{Conflicts of interest: }

No conflicts of interest. 

\subsection*{Funding: }

No funding received. 

\subsection*{Copyright notice: }

Copyright{\small{} }{\footnotesize{}© }2024 Adam Brownstein under
the terms of arXiv.org perpetual, non-exclusive license.

\appendix

\section{Polar decomposition of the Klein-Gordon equation\label{sec:Polar-decomposition-of}}

\noindent The Klein-Gordon equation using the $\eta=(-1,1,1,1)$
metric signature is:
\begin{equation}
\left[\partial_{\mu}\partial^{\mu}-\frac{m^{2}c^{2}}{\hbar^{2}}-\frac{2m}{\hbar^{2}}V\right]\psi=0,\label{eq:1-3}
\end{equation}
where the classical potential $V$ has been rescaled by the factor
$\left(\frac{2m}{\hbar^{2}}\right)^{-1}$. Substituting the polar
form of the wavefunction $\psi=\sqrt{\rho}e^{i\frac{S}{\hbar}}$ into
the Klein-Gordon equation gives:
\begin{alignat}{1}
\left(\partial_{\mu}\partial^{\mu}\sqrt{\rho}\right)e^{i\frac{S}{\hbar}}+\left(\frac{2i}{\hbar}\partial_{\mu}S\partial^{\mu}\sqrt{\rho}\right)e^{i\frac{S}{\hbar}}+\left(\frac{i}{\hbar}\partial_{\mu}^{2}S\right)\sqrt{\rho}e^{i\frac{S}{\hbar}}\ldots\nonumber \\
\ldots+\left(\frac{-1}{\hbar^{2}}\partial_{\mu}S\partial^{\mu}S\right)\sqrt{\rho}e^{i\frac{S}{\hbar}}-\frac{m^{2}c^{2}}{\hbar^{2}}\sqrt{\rho}e^{i\frac{S}{\hbar}}-\frac{2m}{\hbar^{2}}V\sqrt{\rho}e^{i\frac{S}{\hbar}}=0.
\end{alignat}
Multiply by $e^{-i\frac{S}{\hbar}}$ to give:
\begin{equation}
\partial_{\mu}\partial^{\mu}\sqrt{\rho}+\frac{2i}{\hbar}\partial_{\mu}S\partial^{\mu}\sqrt{\rho}+\frac{i}{\hbar}\sqrt{\rho}\partial_{\mu}^{2}S-\frac{1}{\hbar^{2}}\sqrt{\rho}\partial_{\mu}S\partial^{\mu}S-\sqrt{\rho}\frac{m^{2}c^{2}}{\hbar^{2}}-\frac{2}{\hbar^{2}}V\sqrt{\rho}=0.
\end{equation}
Separate the real component:
\begin{equation}
\frac{\sqrt{\rho}\partial_{\mu}S\partial^{\mu}S}{\hbar^{2}}-\partial_{\mu}\partial^{\mu}\sqrt{\rho}+\sqrt{\rho}\frac{m^{2}c^{2}}{\hbar^{2}}+\frac{2m}{\hbar^{2}}V\sqrt{\rho}=0.\label{eq:5-2}
\end{equation}
Multiply by $\frac{\hbar^{2}}{2m\sqrt{\rho}}$:
\begin{equation}
\frac{1}{2m}\partial_{\mu}S\partial^{\mu}S-\frac{\hbar^{2}}{2m}\frac{\partial_{\mu}\partial^{\mu}\sqrt{\rho}}{\sqrt{\rho}}+\frac{1}{2}mc^{2}+V=0.\label{eq:2_a-1}
\end{equation}
This is the relativistic generalization of the quantum Hamilton-Jacobi
equation. To obtain the continuity equation, separate the imaginary
component of the Klein-Gordon equation:
\begin{equation}
\frac{2}{\hbar}\partial_{\mu}S\partial^{\mu}\sqrt{\rho}+\frac{1}{\hbar}\sqrt{\rho}\partial_{\mu}^{2}S=0.\label{eq:7-1}
\end{equation}
Multiply by $\hbar\sqrt{\rho}$:
\begin{align}
0= & 2\partial_{\mu}S\sqrt{\rho}\partial^{\mu}\sqrt{\rho}+\rho\partial_{\mu}^{2}S\\
= & \partial_{\mu}S\partial^{\mu}\rho+\rho\partial_{\mu}^{2}S,\\
= & \partial_{\mu}\left(\rho\partial^{\mu}S\right),\label{eq:2-3}
\end{align}
which is the relativistic generalization of the continuity equation.

\section{Decomposition of the quantum potential \label{sec:Quantum-potential}}

\noindent The quantum potential is defined as:
\begin{equation}
Q\equiv-\frac{\hbar^{2}}{2m}\frac{\partial_{\mu}\partial^{\mu}\sqrt{\rho}}{\sqrt{\rho}}.
\end{equation}
The following steps can be used to decompose the quantum potential
into a form useful for transforming the quantum Cauchy-momentum equations:
\begin{align}
 & \left(-\frac{\hbar^{2}}{2m}\right)^{-1}\rho\partial_{\nu}Q\nonumber \\
 & =\rho\partial_{\nu}\left(\sqrt{\rho}^{-1}\partial_{\mu}\partial^{\mu}\sqrt{\rho}\right)\text{}\\
 & =\sqrt{\rho}\partial_{\nu}\partial_{\mu}\partial^{\mu}\sqrt{\rho}-\partial_{\nu}\sqrt{\rho}\partial_{\mu}\partial^{\mu}\sqrt{\rho}\text{ }\mathcal{\text{\emph{(product rule)}}}\\
 & =\left[\partial_{\mu}\left(\sqrt{\rho}\partial_{\nu}\partial^{\mu}\sqrt{\rho}\right)-\partial_{\mu}\sqrt{\rho}\partial_{\nu}\partial^{\mu}\sqrt{\rho}\right]-\left[\partial_{\mu}\left(\partial_{\nu}\sqrt{\rho}\partial^{\mu}\sqrt{\rho}\right)-\partial_{\nu}\partial_{\mu}\sqrt{\rho}\partial^{\mu}\sqrt{\rho}\right]\text{ \emph{(inverse product rule)}}\\
 & =\left[\partial_{\mu}\left(\sqrt{\rho}\partial_{\nu}\partial^{\mu}\sqrt{\rho}\right)-\partial_{\mu}\left(\partial_{\nu}\sqrt{\rho}\partial^{\mu}\sqrt{\rho}\right)\right]+\left[\partial_{\nu}\partial_{\mu}\sqrt{\rho}\partial^{\mu}\sqrt{\rho}-\partial_{\mu}\sqrt{\rho}\partial_{\nu}\partial^{\mu}\sqrt{\rho}\right]\text{ \emph{(rearrange)}}\\
 & =\partial_{\mu}\left(\sqrt{\rho}\partial_{\nu}\partial^{\mu}\sqrt{\rho}\right)-\partial_{\mu}\left(\partial_{\nu}\sqrt{\rho}\partial^{\mu}\sqrt{\rho}\right)\text{ \emph{(cancellation of second bracket)}}\\
 & =\partial_{\mu}\partial^{\mu}\left(\sqrt{\rho}\partial_{\nu}\sqrt{\rho}\right)-2\partial_{\mu}\left(\partial_{\nu}\sqrt{\rho}\partial^{\mu}\sqrt{\rho}\right)\text{\emph{(inverse product rule)}}\text{ }\\
 & =\frac{1}{2}\partial_{\mu}\partial^{\mu}\partial_{\nu}\rho-2\partial_{\mu}\left(\partial_{\nu}\sqrt{\rho}\partial^{\mu}\sqrt{\rho}\right)\text{\emph{(differentiation)}}.
\end{align}
Restore the constant by multiplying through by $-\frac{\hbar^{2}}{2m}$.
Therefore:
\begin{align}
\rho\partial_{\nu}Q= & -\frac{\hbar^{2}}{4m}\partial_{\mu}\partial^{\mu}\partial_{\nu}\rho+\frac{\hbar^{2}}{m}\partial_{\mu}\left(\partial_{\nu}\sqrt{\rho}\partial^{\mu}\sqrt{\rho}\right)\\
= & -\frac{\hbar^{2}}{4m}\partial_{\mu}\partial^{\mu}\partial_{\nu}\rho+\frac{\hbar^{2}}{m}\partial_{\mu}\left(\rho\frac{\partial_{\nu}\sqrt{\rho}}{\sqrt{\rho}}\frac{\partial^{\mu}\sqrt{\rho}}{\sqrt{\rho}}\right)\\
= & -\frac{\hbar^{2}}{4m}\partial_{\mu}\partial^{\mu}\partial_{\nu}\rho+\frac{\hbar^{2}}{4m}\partial_{\mu}\left(\rho\partial_{\nu}\log\rho\partial^{\mu}\log\rho\right),\label{eq: D11}
\end{align}
as required. A similar derivation works for the non-relativistic case
with result:
\begin{align}
\rho\partial_{j}Q= & -\frac{\hbar^{2}}{4m}\nabla^{2}\partial_{j}\rho+\frac{\hbar^{2}}{4m}\nabla\cdot\left(\rho\partial_{j}\log\rho\nabla\log\rho\right).\label{eq:b13}
\end{align}

\section{\label{sec:Alternative-transformation-of}Alternative transformations
of the quantum potential}

\noindent The quantum potential term can be written in the following
way:
\begin{align}
Q= & -\frac{\hbar^{2}}{2m}\frac{\partial_{\mu}\partial^{\mu}\sqrt{\rho}}{\sqrt{\rho}}\\
= & -\frac{\hbar^{2}}{4m}\frac{\partial_{\mu}\left(\rho^{-\frac{1}{2}}\partial^{\mu}\rho\right)}{\sqrt{\rho}}\\
= & \frac{\hbar^{2}}{8m}\frac{\rho^{-\frac{3}{2}}\partial_{\mu}\rho\partial^{\mu}\rho}{\sqrt{\rho}}-\frac{\hbar^{2}}{4m}\frac{\partial_{\mu}\partial^{\mu}\rho}{\rho}\\
= & \frac{\hbar^{2}}{8m}\partial_{\mu}\log\rho\partial^{\mu}\log\rho-\frac{\hbar^{2}}{4m}\frac{\partial_{\mu}\partial^{\mu}\rho}{\rho}.\label{eq:C4}
\end{align}
It can also be written as:
\begin{align}
Q= & -\frac{\hbar^{2}}{2m}\frac{\partial_{\mu}\partial^{\mu}\sqrt{\rho}}{\sqrt{\rho}}\\
= & -\frac{\hbar^{2}}{2m}\rho^{-\frac{1}{2}}\partial_{\mu}\left(\sqrt{\rho}\frac{1}{\sqrt{\rho}}\partial^{\mu}\sqrt{\rho}\right)\\
= & -\frac{\hbar^{2}}{2m}\frac{\partial_{\mu}\sqrt{\rho}}{\sqrt{\rho}}\frac{\partial^{\mu}\sqrt{\rho}}{\sqrt{\rho}}-\frac{\hbar^{2}}{2m}\partial_{\mu}\left(\frac{1}{\sqrt{\rho}}\partial^{\mu}\sqrt{\rho}\right)\\
= & -\frac{\hbar^{2}}{8m}\partial_{\mu}\log\rho\partial^{\mu}\log\rho-\frac{\hbar^{2}}{4m}\partial_{\mu}\partial^{\mu}\log\rho.\label{eq:C.8}
\end{align}
Adding the first and second forms then dividing by two gives a third
form:
\begin{align}
Q= & -\frac{\hbar^{2}}{8m}\partial_{\mu}\partial^{\mu}\log\rho-\frac{\hbar^{2}}{8m}\frac{\partial_{\mu}\partial^{\mu}\rho}{\rho}.\label{eq:126}
\end{align}
The non-relativistic version of these equations can be derived from
$Q\equiv-\frac{\hbar^{2}}{2m}\frac{\nabla^{2}\sqrt{\rho}}{\sqrt{\rho}}$
following identical steps, with the results:
\begin{align}
Q= & \frac{\hbar^{2}}{8m}\nabla\log\rho\cdot\nabla\log\rho-\frac{\hbar^{2}}{4m}\frac{\nabla^{2}\rho}{\rho}\label{eq:c10}\\
Q= & -\frac{\hbar^{2}}{8m}\nabla\log\rho\cdot\nabla\log\rho-\frac{\hbar^{2}}{4m}\nabla^{2}\log\rho\\
Q= & -\frac{\hbar^{2}}{8m}\nabla^{2}\log\rho-\frac{\hbar^{2}}{8m}\frac{\nabla^{2}\rho}{\rho}.\label{eq:C.12}
\end{align}

\section{\label{sec:Stochastic-transport}Stochastic Lagrangian derivative}

\noindent In this section we derive the stochastic Lagrangian derivative,
which can be found using Taylor expansion methods. The forward-causal
stochastic Lagrangian derivative of a function $f$ is defined as:
\begin{equation}
\frac{Df(x,\tau)}{D\tau}=E\left[\frac{f(x,\tau)-f(x-dx,\tau-d\tau)}{d\tau}\right],
\end{equation}
where the expectation is over stochastic trajectories $dx$ leading
to the coordinate $(x,\tau)$. It represents the update of the quantity
$f(x-dx,\tau-d\tau)$ along a range of particle trajectories, where
the result is averaged at the single location $(x,\tau)$. The function
$f(x-dx,\delta\tau)$ can be expanded to second order using the Taylor
series:
\begin{align}
\frac{Df(x,\tau)}{D\tau}\equiv & E\left[\frac{f(x,\tau)-f(x-dx,\tau-d\tau)}{d\tau}\right]\\
\approx & E\left[\frac{\partial f(x,\tau)}{\partial\tau}+\frac{\partial f(x,\tau)}{\partial x^{\mu}}\frac{dx^{\mu}}{d\tau}-\frac{1}{2}\frac{\partial^{2}f(x,\tau)}{\partial x^{\mu}\partial x^{\nu}}\frac{dx^{\mu}dx^{\nu}}{d\tau}\right]\label{eq:d3}\\
= & \frac{\partial f(x,\tau)}{\partial\tau}+\frac{dx^{\mu}}{d\tau}\frac{\partial f(x,\tau)}{\partial x^{\mu}}-\frac{1}{2}k\partial_{\mu}\partial^{\mu}f(x,\tau)\\
= & \left[\frac{\partial}{\partial\tau}+v^{\mu}\frac{\partial}{\partial x^{\mu}}-\frac{1}{2}k\partial_{\mu}\partial^{\mu}\right]f(x,\tau),
\end{align}
where $k$ is a diffusion constant (e.g. equal to $\frac{\hbar}{m}$
in previous sections), $v^{\mu}$ is the drift component of the particle's
four-velocity (e.g. equal to $\frac{\partial^{\mu}S'}{m}$ in previous
sections) and $E\left[dx^{\mu}dx^{\mu}\right]=k\delta^{\mu\nu}d\tau$
has been used in {[}Eq. \ref{eq:d3}{]}. Therefore the forward-causal
stochastic Lagrangian derivative is defined as:
\begin{align}
\frac{D}{D\tau}\equiv & \frac{\partial}{\partial\tau}+v^{\mu}\frac{\partial}{\partial x^{\mu}}-\frac{k}{2}\partial_{\mu}\partial^{\mu}.
\end{align}
The fact that this is the correct definition for the forward-causal
stochastic Lagrangian derivative can also be seen in the standard
forward-causal Fokker-Planck equation:
\begin{equation}
\frac{\partial\rho}{\partial t}+\nabla\cdot\left(\rho{\bf v}'\right)=\frac{k}{2}\nabla^{2}\rho,
\end{equation}
which can be written in terms of the stochastic Lagrangian derivative
as:
\begin{equation}
\frac{D\rho}{D\tau}+\rho\nabla\cdot{\bf v}'=0.
\end{equation}
which indicates that the minus sign on the $\frac{k}{2}\nabla^{2}$
term is correct. By comparison, the retrocausal version of the stochastic
Lagrangian derivative has a positive sign on the $\frac{k}{2}\nabla^{2}$
term, and is defined as:
\begin{align}
\frac{D_{*}}{D\tau}\equiv & \frac{\partial}{\partial\tau}+v^{\mu}\frac{\partial}{\partial x^{\mu}}+\frac{k}{2}\partial_{\mu}\partial^{\mu}.
\end{align}

\section{\label{sec:Quantum-Hamilton-Jacobi-equation}Quantum Hamilton-Jacobi
equation}

\subsection{Relativistic case: }

It can be shown by substituting the first form of the quantum potential
\emph{{[}Eq. \ref{eq:C4}{]}:}
\begin{equation}
Q=\frac{\hbar^{2}}{8m}\partial_{\mu}\log\rho\partial^{\mu}\log\rho-\frac{\hbar^{2}}{4m}\frac{\partial_{\mu}\partial^{\mu}\rho}{\rho},\label{eq:E1}
\end{equation}
and using the stochastic transform:
\begin{equation}
S'=S+\frac{\hbar}{2}\log\rho,
\end{equation}
that the quantum Hamilton-Jacobi equation becomes:
\begin{equation}
\frac{1}{2m}\partial_{\mu}S'\partial^{\mu}S'+\frac{\hbar}{2m}\partial_{\mu}\partial^{\mu}S'+2Q+V+\frac{1}{2}m=0.
\end{equation}
We will give a brief proof of this result. The quantum Hamilton-Jacobi
equation is:
\begin{align}
\frac{1}{2m}\partial_{\mu}S\partial^{\mu}S+Q+V+\frac{1}{2}m=0.
\end{align}
Substitute the form of the quantum potential {[}Eq. \ref{eq:E1}{]}
to give:
\begin{align}
\frac{1}{2m}\partial_{\mu}S\partial^{\mu}S+\frac{\hbar^{2}}{8m}\partial_{\mu}\log\rho\partial^{\mu}\log\rho-\frac{\hbar^{2}}{4m}\frac{\partial_{\mu}\partial^{\mu}\rho}{\rho}+V+\frac{1}{2}m=0,
\end{align}
therefore by using the forward-causal stochastic transform of the
phase $S'=S+\frac{\hbar}{2}\log\rho$:
\begin{align}
\frac{1}{2m}\partial_{\mu}S'\partial^{\mu}S'-\frac{\hbar}{2m}\partial_{\mu}S\partial^{\mu}\log\rho-\frac{\hbar^{2}}{4m}\frac{\partial_{\mu}\partial^{\mu}\rho}{\rho}+V+\frac{1}{2}m=0.
\end{align}
Therefore using the continuity equation $\partial_{\mu}S\partial^{\mu}\log\rho=-\partial_{\mu}\partial^{\mu}S$:
\begin{align}
\frac{1}{2m}\partial_{\mu}S'\partial^{\mu}S'+\frac{\hbar}{2m}\partial_{\mu}\partial^{\mu}S-\frac{\hbar^{2}}{4m}\frac{\partial_{\mu}\partial^{\mu}\rho}{\rho}+V+\frac{1}{2}m=0,
\end{align}
and using the forward-causal stochastic transformation of the phase
on the term $\frac{\hbar}{2m}\partial_{\mu}\partial^{\mu}S$: 
\begin{align}
\frac{1}{2m}\partial_{\mu}S'\partial^{\mu}S'+\frac{\hbar}{2m}\partial_{\mu}\partial^{\mu}S'-\frac{\hbar^{2}}{4m}\partial_{\mu}\partial^{\mu}\log\rho-\frac{\hbar^{2}}{4m}\frac{\partial_{\mu}\partial^{\mu}\rho}{\rho}+V+\frac{1}{2}m=0.
\end{align}
Now the remaining terms $-\frac{\hbar^{2}}{4m}\partial_{\mu}\partial^{\mu}\log\rho-\frac{\hbar^{2}}{4m}\frac{\partial_{\mu}\partial^{\mu}\rho}{\rho}=2Q$
can be identified to be the third-form of the quantum potential {[}Eq.
\ref{eq:C.8}{]}. Therefore:
\begin{align}
\frac{1}{2m}\partial_{\mu}S'\partial^{\mu}S'+\frac{\hbar}{2m}\partial_{\mu}\partial^{\mu}S'+2Q+V+\frac{1}{2}m=0,
\end{align}
as required: We can also add $\frac{\partial S'}{\partial\tau}=0$
to the equation to give: 
\begin{align}
\frac{\partial S'}{\partial\tau}+\frac{1}{2m}\partial_{\mu}S'\partial^{\mu}S'+\frac{\hbar}{2m}\partial_{\mu}\partial^{\mu}S'+2Q+V+\frac{1}{2}m=0.\label{eq:E11}
\end{align}

\subsection{Gradient of the quantum Hamilton-Jacobi equation: }

The gradient of the quantum Hamilton-Jacobi equation {[}Eq. \ref{eq:E11}{]}
is: 
\begin{align}
\frac{\partial\partial_{\nu}S'}{\partial\tau}+\frac{1}{m}\partial_{\mu}S'\partial^{\mu}\partial_{\nu}S'+\frac{\hbar}{2m}\partial_{\mu}\partial^{\mu}\partial_{\nu}S'+\partial_{\nu}V+2\partial_{\nu}Q= & 0.\label{eq:e15}
\end{align}
Using the retrocausal stochastic Lagrangian derivative: 
\begin{align}
\frac{D_{*}}{D\tau}\equiv\frac{\partial}{\partial\tau}+\frac{1}{m}\partial_{\mu}S'\partial_{\mu}+\frac{\hbar}{2m}\partial_{\mu}\partial^{\mu},
\end{align}
this can be written as: 
\begin{align}
\frac{D_{*}\partial_{\nu}S'}{D\tau}+\partial_{\nu}V+2\partial_{\nu}Q= & 0.\label{eq:e16-1}
\end{align}
Using the forward-causal stochastic Lagrangian derivative: 
\begin{align}
\frac{D}{D\tau}\equiv\frac{\partial}{\partial\tau}+\frac{1}{m}\partial_{\mu}S'\partial_{\mu}-\frac{\hbar}{2m}\partial_{\mu}\partial^{\mu},
\end{align}
it can be written as:
\begin{align}
\frac{D\partial_{\nu}S'}{D\tau}+\partial_{\nu}V+2\partial_{\nu}Q+\frac{\hbar}{m}\partial_{\mu}\partial^{\mu}\partial_{\nu}S'= & 0.\label{eq:e16}
\end{align}

\subsection{Non-relativistic case: }

It can be shown in an analogous way to the relativistic case that
by substituting the form of the quantum potential {[}Eq. \ref{eq:c10}{]}:
\begin{equation}
Q=\frac{\hbar^{2}}{8m}\nabla\log\rho\cdot\nabla\log\rho-\frac{\hbar^{2}}{4m}\frac{\nabla^{2}\rho}{\rho},
\end{equation}
and using the stochastic transform:
\begin{equation}
S=S'+\frac{\hbar}{2}\log\rho,
\end{equation}
and continuity equation:
\begin{equation}
\frac{\partial\rho}{\partial t}+\nabla\cdot\left(\rho\nabla S\right)=0,
\end{equation}
that the quantum Hamilton-Jacobi equation becomes:
\begin{equation}
\frac{\partial S'}{\partial t}+\frac{1}{2m}\nabla S'\cdot\nabla S'+\frac{\hbar}{2m}\nabla^{2}S'+V+2Q=0.
\end{equation}
In terms of the retrocausal stochastic Lagrangian derivative, the
gradient of the Hamilton-Jacobi equation is equal to:
\begin{equation}
\frac{D_{*}\nabla S'}{Dt}+\nabla V+2\nabla Q=0.
\end{equation}
In terms of the forward-causal stochastic Lagrangian derivative, the
gradient it is equal to:
\begin{equation}
\frac{D\nabla S'}{Dt}+\nabla V+2\nabla Q+\frac{\hbar}{m}\nabla^{2}\nabla S'=0.
\end{equation}

\section{Non-relativistic Cauchy-momentum equations}

\subsection{\label{subsec:Non-relativistic-case}Cauchy-momentum equations}

\noindent Here we will discuss whether the stochastic transformation
can be used to remove the quantum potential term in the non-relativistic
case. We will first derive the Cauchy-momentum equation for the non-relativistic
Schrödinger equation. To do this, find $\frac{\partial\rho\partial_{j}S}{\partial t}$:
\begin{align}
\frac{\partial\rho\partial_{j}S}{\partial t}= & \frac{\partial\rho}{\partial t}\partial_{j}S+\rho\frac{\partial\partial_{j}S}{\partial t}\label{eq:3.23}
\end{align}
The continuity equation {[}Eq. \ref{eq:2-2}{]} implies:
\begin{equation}
\frac{\partial\rho}{\partial t}\partial_{j}S=-\partial_{j}S\nabla\cdot\left(\rho\frac{\nabla S}{m}\right).
\end{equation}
Meanwhile, the quantity $\rho\frac{\partial\partial_{j}S}{\partial t}$
can be found by taking the derivative of the quantum Hamilton-Jacobi
equation {[}Eq. \ref{eq:1-2}{]} and then multiplying through by $\rho$:
\begin{align}
\rho\frac{\partial\partial_{j}S}{\partial t}=-\frac{1}{m}\rho\nabla S\cdot\nabla\partial_{j}S-\rho\partial_{j}V-\rho\partial_{j}Q.
\end{align}
Substituting both of these expressions into {[}Eq. \ref{eq:3.23}{]}
gives:
\begin{align}
\frac{\partial\rho\partial_{j}S}{\partial t}=-\frac{1}{m}\partial_{j}S\nabla\cdot\left(\rho\nabla S\right)-\frac{1}{m}\rho S\cdot\nabla\partial_{j}S-\rho\partial_{j}V-\rho\partial_{j}Q=0,
\end{align}
which simplifies to the Cauchy-momentum equation:
\begin{align}
\frac{\partial\rho\partial_{j}S}{\partial t}+\frac{1}{m}\nabla\cdot\left(\rho\partial_{j}S\nabla S\right)+\rho\partial_{j}V+\rho\partial_{j}Q=0.
\end{align}
The next step is to decompose the quantum potential term and absorb
it into the kinetic term. Appendix \ref{sec:Quantum-potential} {[}Eq.
\ref{eq:b13}{]} shows that the quantum potential term $\rho\partial_{j}Q$
can be expressed as:
\begin{equation}
\rho\partial_{j}Q=-\frac{\hbar^{2}}{4m}\nabla^{2}\partial_{j}\rho+\frac{\hbar^{2}}{4m}\nabla\cdot\left(\rho\partial_{j}\log\rho\nabla\log\rho\right),
\end{equation}
therefore:
\begin{align}
\frac{\partial\rho\partial_{j}S}{\partial t}+\frac{1}{m}\nabla\cdot\left(\rho\partial_{j}S\nabla S\right)+\rho\partial_{j}V+\frac{\hbar^{2}}{4m}\nabla\cdot\left(\rho\partial_{j}\log\rho\nabla\log\rho\right)-\frac{\hbar^{2}}{4m}\nabla^{2}\partial_{j}\rho= & 0.
\end{align}
Collecting like terms gives:
\begin{align}
\frac{\partial\rho\partial_{j}S}{\partial t}+\frac{1}{m}\nabla\cdot\left(\rho\partial_{j}S\nabla S+\frac{\hbar^{2}}{4}\rho\partial_{j}\log\rho\nabla\log\rho\right)+\rho\partial_{j}V-\frac{\hbar^{2}}{4m}\nabla^{2}\partial_{j}\rho & =0.
\end{align}
Adding and subtracting the cross terms $\partial_{j}S\nabla\log\rho$
and $\partial_{j}\log\rho\nabla S$ gives:
\begin{align}
 & \frac{\partial\rho\partial_{j}S}{\partial t}+\frac{1}{m}\nabla\cdot\left[\rho\left(\partial_{j}S+\frac{\hbar}{2}\partial_{j}\log\rho\right)\left(\nabla S+\frac{\hbar}{2}\nabla\log\rho\right)\right]\ldots\nonumber \\
 & \ldots-\frac{\hbar}{2m}\nabla\cdot\left(\rho\partial_{j}\log\rho\nabla S\right)-\frac{\hbar}{2m}\nabla\cdot\left(\rho\partial_{j}S\nabla\log\rho\right)+\rho\partial_{j}V-\frac{\hbar^{2}}{4m}\nabla^{2}\partial_{j}\rho=0.
\end{align}
After performing the stochastic transformation $S'=S+\frac{\hbar}{2}\log\rho$
we get:
\begin{align}
\frac{\partial\rho\partial_{j}S}{\partial t}+\frac{1}{m}\nabla\cdot\left(\rho\partial_{j}S'\nabla S'\right)+\rho\partial_{j}V-\frac{\hbar}{2m}\nabla\cdot\left(\partial_{j}\rho\nabla S\right)-\frac{\hbar}{2m}\nabla\cdot\left(\nabla\rho\partial_{j}S\right)= & \frac{\hbar^{2}}{4m}\nabla^{2}\partial_{j}\rho.\label{eq:3.30}
\end{align}
Now the objective is to cancel the cross terms. To do so, recognize
that the stochastic transform needs to also be performed on $\frac{\partial\rho\partial_{j}S}{\partial t}$:
\begin{align}
\frac{\partial\rho\partial_{j}S}{\partial t}= & \frac{\partial\rho\partial_{j}S'}{\partial t}-\frac{\hbar}{2}\frac{\partial\rho\partial_{j}\log\rho}{\partial t}\\
= & \frac{\partial\rho\partial_{j}S'}{\partial t}-\frac{\hbar}{2}\partial_{j}\frac{\partial\rho}{\partial t}\\
= & \frac{\partial\rho\partial_{j}S'}{\partial t}+\frac{\hbar}{2m}\partial_{j}\nabla\cdot\left(\rho\nabla S\right),\label{eq:3.33}
\end{align}
where we have used the continuity equation to write $-\frac{\hbar}{2}\partial_{j}\frac{\partial\rho}{\partial t}=\frac{\hbar}{2m}\partial_{j}\nabla\cdot\left(\rho\nabla S\right)$.
The resulting $\frac{\hbar}{2m}\partial_{j}\nabla\cdot\left(\rho\nabla S\right)$
term can partially cancel the cross terms as follows. Substitute {[}Eq.
\ref{eq:3.33}{]} into {[}Eq. \ref{eq:3.30}{]}: 
\begin{align}
\frac{\partial\rho\partial_{j}S'}{\partial t}+\frac{1}{m}\nabla\cdot\left(\rho\partial_{j}S'\nabla S'\right)+\rho\partial_{j}V+\ldots\ \ \ \ \ \ \ \ \ \ \ \ \ \ \ \ \ \ \ \ \ \ \ \ \ \ \ \ \ \ \ \ \ \ \ \ \nonumber \\
\ldots+\left[\frac{\hbar}{2m}\partial_{j}\nabla\cdot\left(\rho\nabla S\right)-\frac{\hbar}{2m}\nabla\cdot\left(\partial_{j}\rho\nabla S\right)-\frac{\hbar}{2m}\nabla\cdot\left(\nabla\rho\partial_{j}S\right)\right]= & \frac{\hbar^{2}}{4m}\nabla^{2}\partial_{j}\rho.
\end{align}
Simplify the factor in the square bracket:
\begin{align}
 & \frac{\hbar}{2m}\partial_{j}\nabla\cdot\left(\rho\nabla S\right)-\frac{\hbar}{2m}\nabla\cdot\left(\partial_{j}\rho\nabla S\right)-\frac{\hbar}{2m}\nabla\cdot\left(\nabla\rho\partial_{j}S\right)\nonumber \\
 & =\frac{\hbar}{2m}\partial_{j}\nabla\cdot\left(\rho\nabla S\right)-\left[\frac{\hbar}{2m}\partial_{j}\nabla\cdot\left(\rho\nabla S\right)-\frac{\hbar}{2m}\nabla\cdot\left(\rho\nabla\partial_{j}S\right)\right]-\left[\frac{\hbar}{2m}\nabla^{2}\left(\rho\partial_{j}S\right)-\frac{\hbar}{2m}\nabla\cdot\left(\rho\nabla\partial_{j}S\right)\right]\\
 & =-\frac{\hbar}{2m}\nabla^{2}\left(\rho\partial_{j}S\right)+\frac{\hbar}{m}\nabla\cdot\left(\rho\nabla\partial_{j}S\right).
\end{align}
Therefore:
\begin{align}
\frac{\partial\rho\partial_{j}S}{\partial t}+\frac{1}{m}\nabla\cdot\left(\rho\partial_{j}S'\nabla S'\right)+\rho\partial_{j}V+\frac{\hbar}{m}\nabla\left(\rho\nabla\partial_{j}S\right)= & \frac{\hbar}{2m}\nabla^{2}\left(\rho\partial_{j}S\right)+\frac{\hbar^{2}}{4m}\nabla^{2}\partial_{j}\rho.
\end{align}
Combine the two terms on the right hand side using the stochastic
transformation:
\begin{align}
\frac{\partial\rho\partial_{j}S'}{\partial t}+\frac{1}{m}\nabla\cdot\left(\rho\partial_{j}S'\nabla S'\right)+\rho\partial_{j}V+\frac{\hbar}{m}\nabla\cdot\left(\rho\nabla\partial_{j}S\right)= & \frac{\hbar}{2m}\nabla^{2}\left(\rho\partial_{j}S'\right),
\end{align}
so we get a similar transformed Cauchy-momentum equation as in the
relativistic case. However, we must ask whether the term $\nabla\cdot\left(\rho\nabla\partial_{j}S\right)$
equals zero as it did for the relativistic Klein-Gordon equations,
where it was seen that $\partial_{\mu}\left(\rho\partial^{\mu}\partial_{j}S\right)=0$.
This is crucial because if the term $\nabla\cdot\left(\rho\nabla\partial_{j}S\right)$
is zero, then the Cauchy-momentum equations are classical under the
stochastic transform. But we know via Bell's theorem that the non-relativistic
case cannot allow a local hidden variable theory, and so it must be
the case that $\nabla\cdot\left(\rho\nabla\partial_{j}S\right)$ is
not zero for all coordinates. 

The suspected non-vanishing of this quantity therefore provides an
insight into the mechanism responsible for non-locality. It is this
term which replaces the quantum potential as the source of non-locality.
Note that the problem of non-locality is evaded in the Klein-Gordon
system because it is formulated as a stochastic diffusion in spacetime
coordinates and hence has retrocausal effects, which provide a loophole
against Bell's theorem. But this is not possible for the non-relativistic
Schrodinger system, which is formulated as a stochastic diffusion
in position coordinates only. Furthermore, for the Klein-Gordon system,
the condition $\partial_{\mu}\left(\rho\partial^{\mu}\partial_{\nu}S\right)=0$
takes on the form of a conservation law. In the non-relativistic case
the equivalent conservation law would be: 
\begin{equation}
\frac{\partial\rho\partial_{j}S}{\partial t}+\nabla\cdot\left(\rho\nabla\partial_{j}S\right)=0,
\end{equation}
and $\nabla\cdot\left(\rho\nabla\partial_{j}S\right)=0$ would then
imply that $\frac{\partial\rho\partial_{j}S}{\partial t}$ vanishes,
which is not true in general. For example, the term $\frac{\partial\rho\partial_{j}S}{\partial t}$
is part of the deterministic Cauchy-momentum equation: 
\begin{align}
\frac{\partial\rho\partial_{j}S}{\partial t}+\frac{1}{m}\nabla\cdot\left(\rho\partial_{j}S\nabla S\right)+\rho\partial_{j}V+\rho\partial_{j}Q= & 0,
\end{align}
and the condition $\frac{\partial\rho\partial_{j}S}{\partial t}=0$
is only valid for states with stationary momentum.

\section{\label{sec:Consistency-condition}Consistency lemma}

\noindent In this section, we will prove the condition:
\begin{equation}
-2\partial_{\nu}Q-\frac{\hbar}{m}\partial_{\mu}\partial^{\mu}\partial_{\nu}S'=\frac{\hbar}{m}\partial_{\mu}\rho\partial^{\mu}\partial_{\nu}S',\label{eq:G1}
\end{equation}
which is discussed in section \ref{subsec:Consistency-condition}.
Substitute the form of the quantum potential {[}Eq. \ref{eq:C.8}{]}
into the left-hand side of {[}Eq. \ref{eq:G1}{]}:
\begin{equation}
Q=-\frac{\hbar^{2}}{8m}\partial_{\mu}\log\rho\partial^{\mu}\log\rho-\frac{\hbar^{2}}{4m}\partial_{\mu}\partial^{\mu}\log\rho,
\end{equation}
therefore:
\begin{align}
-2Q-\frac{\hbar}{m}\partial_{\mu}\partial^{\mu}S'= & \left[\frac{\hbar^{2}}{4m}\partial_{\mu}\log\rho\partial^{\mu}\log\rho+\frac{\hbar^{2}}{2m}\partial_{\mu}\partial^{\mu}\log\rho\right]-\frac{\hbar}{m}\partial_{\mu}\partial^{\mu}S-\frac{\hbar}{2m}\partial_{\mu}\partial^{\mu}\log\rho\\
= & \frac{\hbar^{2}}{4m}\partial_{\mu}\log\rho\partial^{\mu}\log\rho-\frac{\hbar}{m}\partial_{\mu}\partial^{\mu}S.\label{eq:7.18}
\end{align}
Now by taking the gradient of {[}Eq. \ref{eq:7.18}{]} and multiplying
through by $\rho$, we have: 
\begin{align}
-2\rho\partial_{\nu}Q-\frac{\hbar}{m}\rho\partial_{\mu}\partial^{\mu}\partial_{\nu}S'= & \frac{\hbar^{2}}{2m}\partial_{\mu}\rho\partial^{\mu}\partial_{\nu}\log\rho-\frac{\hbar}{m}\rho\partial_{\mu}\partial^{\mu}\partial_{\nu}S.\\
= & \frac{\hbar^{2}}{2m}\partial_{\mu}\rho\partial^{\mu}\partial_{\nu}\log\rho+\frac{\hbar}{m}\partial_{\mu}\rho\partial^{\mu}\partial_{\nu}S\label{eq:61}\\
= & \frac{\hbar}{m}\partial_{\mu}\rho\partial^{\mu}\partial_{\nu}S'.\label{eq:62}
\end{align}
where we have used the rank-2 current conservation condition $\partial_{\mu}\left(\rho\partial^{\mu}\partial_{\nu}S\right)=0$
to write $\rho\partial_{\mu}\partial^{\mu}\partial_{\nu}S=-\partial_{\mu}\rho\partial^{\mu}\partial_{\nu}S$
in {[}Eq. \ref{eq:61}{]}, and have used the stochastic transform
$S'=S+\frac{\hbar}{2}\log\rho$ to simplify. 

\section{phase space approach to the equivariance relationship\label{sec:phase-space-approach}}

\noindent The de Broglie-Bohm interpretation relies upon the equivariance
relationship, which ensures that a statistical ensemble of particles
have a distribution which match the Born rule probabilities of quantum
mechanics. We show that the equivariance relationship can derived
via two methods, firstly through an argument that the spatial derivative
of the velocity of a point-like particle is zero, and secondly from
the approach of Liouville's theorem in phase space. The second phase
space approach is not commonly discussed in the literature, however
it provides a particular clear formulation of the problem. 

\subsection{Velocity approach}

We assume that a de Broglie-Bohm particle configuration has coordinates
${\bf X}^{a}(t)$. Therefore, the particle configuration is represented
by a localised delta function $\delta({\bf x}-{\bf X}_{t}^{a})$ which
has spatial support at the actual location of the particle configuration
in configuration space. Since the delta function follows along with
the particle configuration, this function is constant in the Lagrangian
reference frame. Therefore: 
\begin{equation}
\frac{d\delta(x-{\bf X}_{t}^{a})}{dt}=0,\label{eq:g.2}
\end{equation}
where the deterministic Lagrangian derivative is defined as: 
\begin{equation}
\frac{d}{dt}=\frac{\partial}{\partial t}+{\bf v}^{a}\cdot\nabla.
\end{equation}
Therefore by expanding the Lagrangian derivative of {[}Eq. \ref{eq:g.2}{]}
we have: 
\begin{equation}
\frac{\partial\delta({\bf x}-{\bf X}_{t}^{a})}{\partial t}+{\bf v}^{a}\cdot\nabla\delta({\bf x}-{\bf X}^{a}(t))=0.\label{eq:H3}
\end{equation}
Since the velocity ${\bf v}^{a}=\frac{d{\bf X}^{a}(t)}{dt}$ is explicitly
a function of time only, its derivative with respect to spatial coordinates
vanishes. Therefore an additional piece $\delta({\bf x}-{\bf X}_{t}^{a})\nabla\cdot{\bf v}^{a}=0$
can be added to {[}Eq. \ref{eq:H3}{]} which results in: 
\begin{align}
0= & \frac{\partial\delta({\bf x}-{\bf X}_{t}^{a})}{\partial t}+{\bf v}^{a}\cdot\nabla\delta({\bf x}-{\bf X}_{t}^{a})+\delta({\bf x}-{\bf X}_{t}^{a})\nabla\cdot{\bf v}^{a}\\
= & \frac{\partial\delta({\bf x}-{\bf X}_{t}^{a})}{\partial t}+\nabla\cdot\left(\delta({\bf x}-{\bf X}_{t}^{a}){\bf v}^{a}\right).
\end{align}
We assume that the particle configuration has the same velocity when
located at the same point in configuration space. Therefore we write
${\bf v}^{a}\delta({\bf x}-{\bf X}_{t}^{a})={\bf v}({\bf x})\delta({\bf x}-{\bf X}_{t}^{a})$,
where ${\bf v}={\bf v}({\bf x})$ is a function of ${\bf x}$, i.e.
it is now a velocity field. Consequently:
\begin{equation}
\frac{\partial\delta({\bf x}-{\bf X}_{t}^{a})}{\partial t}+\nabla\cdot\left(\delta({\bf x}-{\bf X}_{t}^{a}){\bf v}\right)=0
\end{equation}
Averaging over particles in the statistical ensemble and defining
the probability density as:
\begin{equation}
\rho=\lim_{N\rightarrow\infty}\frac{1}{N}\sum_{a=1}^{N}\delta({\bf x}-{\bf X}_{t}^{a}),
\end{equation}
then gives:
\begin{align}
\frac{\partial\lim_{N\rightarrow\infty}\frac{1}{N}\sum_{a=1}^{N}\delta({\bf x}-{\bf X}_{t}^{a})}{\partial t}+\nabla\cdot\left(\lim_{N\rightarrow\infty}\frac{1}{N}\sum_{a=1}^{N}\delta({\bf x}-{\bf X}_{t}^{a}){\bf v}\right)= & 0,
\end{align}
or equivalently:
\begin{align}
\frac{\partial\rho}{\partial t}+\nabla\cdot\left(\rho{\bf v}\right)= & 0.
\end{align}
By choosing ${\bf v}=\nabla S$ this equation becomes identical to
the continuity equation of quantum mechanics. Therefore, the distribution
of particles in the ensemble $\rho=\lim_{N\rightarrow\infty}\frac{1}{N}\sum_{a=1}^{N}\delta({\bf x}-{\bf X}_{t}^{a})$
must match the quantum probability distribution for all subsequent
times if the distributions are initially equal. Therefore the particles
satisfy the equivariance property for their positions. 

\subsection{Equivariance in the phase space approach\label{subsec:Equivariance-in-the}}

The continuity equation for the particle configuration can also be
obtained using a phase space approach. We assume a de Broglie-Bohm
particle configuration has position coordinates ${\bf X}^{a}(t)$
and momentum coordinates ${\bf P}^{a}(t)=m\frac{d{\bf X}(t)}{dt}$
in phase space. Therefore, the particles are represented by a localised
point in phase space $\delta({\bf x}-{\bf X}_{t}^{a})\delta({\bf p}-{\bf P}_{t}^{a})$,
which is a delta function. Since the delta function follows along
with the location of the particle, this function is constant in the
Lagrangian reference frame. Therefore, the Lagrangian derivative of
the delta function is zero:

\begin{equation}
\frac{d\delta({\bf x}-{\bf X}_{t}^{a})\delta({\bf p}-{\bf P}_{t}^{a})}{dt}=0.
\end{equation}
The deterministic Lagrangian derivative for a phase space object is
defined as:
\begin{equation}
\frac{d}{dt}=\frac{\partial}{\partial t}+{\bf \dot{x}}^{a}\cdot\nabla_{{\bf x}}+\dot{{\bf p}}^{a}\cdot\nabla_{{\bf p}},
\end{equation}
where ${\bf \dot{x}}^{a}\equiv\frac{d{\bf X}^{a}(t)}{dt}$ and $\dot{{\bf p}}^{a}\equiv\frac{d{\bf P}^{a}(t)}{dt}$
respectively. Therefore we have:
\begin{equation}
\frac{\partial\delta({\bf x}-{\bf X}_{t}^{a})\delta({\bf p}-{\bf P}_{t}^{a})}{\partial t}+{\bf \dot{x}}^{a}\cdot\nabla_{{\bf x}}\delta({\bf x}-{\bf X}_{t}^{a})\delta({\bf p}-{\bf P}_{t}^{a})+\dot{{\bf p}}^{a}\cdot\nabla_{{\bf p}}\delta({\bf x}-{\bf X}_{t}^{a})\delta({\bf p}-{\bf P}_{t}^{a})=0.\label{eq:G12}
\end{equation}
We assume that the particle configurations have the same velocity
and momenta when located at the same point in phase space. Therefore:
\begin{equation}
{\bf \dot{x}}^{a}\cdot\nabla_{{\bf x}}\delta({\bf x}-{\bf X}_{t}^{a})\delta({\bf p}-{\bf P}_{t}^{a})=\dot{{\bf x}}\cdot\nabla_{{\bf x}}\delta({\bf x}-{\bf X}_{t}^{a})\delta({\bf p}-{\bf P}_{t}^{a}),\label{eq:g13}
\end{equation}
and: 

\begin{equation}
\dot{{\bf p}}^{a}\cdot\nabla_{{\bf p}}\delta({\bf x}-{\bf X}_{t}^{a})\delta({\bf p}-{\bf P}_{t}^{a})=\dot{{\bf p}}\cdot\nabla_{{\bf p}}\delta({\bf x}-{\bf X}_{t}^{a})\delta({\bf p}-{\bf P}_{t}^{a}),\label{eq:g14}
\end{equation}
which is ensured due to the delta functions. The quantities $\dot{{\bf x}}\equiv\dot{{\bf x}}({\bf x},{\bf p})$
and $\dot{{\bf p}}\equiv\dot{{\bf p}}({\bf x},{\bf p})$ are now a
velocity field and momentum field which are functions of the phase
space coordinates. Using {[}Eq. \ref{eq:g13}{]} and {[}Eq. \ref{eq:g14}{]},
{[}Eq. \ref{eq:G12}{]} can be written as: 
\begin{equation}
\frac{\partial\delta({\bf x}-{\bf X}_{t}^{a})\delta({\bf p}-{\bf P}_{t}^{a})}{\partial t}+{\bf \dot{x}}\cdot\nabla_{{\bf x}}\delta({\bf x}-{\bf X}_{t}^{a})\delta({\bf p}-{\bf P}_{t}^{a})+\dot{{\bf p}}\cdot\nabla_{{\bf p}}\delta({\bf x}-{\bf X}_{t}^{a})\delta({\bf p}-{\bf P}_{t}^{a})=0.
\end{equation}
This equation is equivalent to Liouville's theorem, as can be shown
by averaging over particle configurations of the statistical ensemble.
Let the phase space probability density $\rho({\bf x},{\bf p},t)$
be defined as:
\begin{equation}
\rho({\bf x},{\bf p},t)=\lim_{N\rightarrow\infty}\frac{1}{N}\sum_{a=1}^{N}\delta({\bf x}-{\bf X}_{t}^{a})\delta({\bf p}-{\bf P}_{t}^{a}),
\end{equation}
therefore:
\begin{align}
\frac{\partial\lim_{N\rightarrow\infty}\frac{1}{N}\sum_{a=1}^{N}\delta({\bf x}-{\bf X}_{t}^{a})\delta({\bf p}-{\bf P}_{t}^{a})}{\partial t}+\dot{{\bf x}}\cdot\nabla_{{\bf x}}\lim_{N\rightarrow\infty}\frac{1}{N}\sum_{a=1}^{N}\delta({\bf x}-{\bf X}_{t}^{a})\delta({\bf p}-{\bf P}_{t}^{a})\ldots\nonumber \\
\ldots+\dot{{\bf p}}\cdot\nabla_{{\bf p}}\lim_{N\rightarrow\infty}\frac{1}{N}\sum_{a=1}^{N}\delta({\bf x}-{\bf X}_{t}^{a})\delta({\bf p}-{\bf P}_{t}^{a}) & =0,\label{eq:G14}
\end{align}
or equivalently: 
\begin{equation}
\frac{\partial\rho}{\partial t}+\dot{{\bf x}}\cdot\nabla_{{\bf x}}\rho+\dot{{\bf p}}\cdot\nabla_{{\bf p}}\rho=0,\label{eq:G.15}
\end{equation}
which is Liouville's theorem. Now it remains to show that the continuity
equation is also satisfied. To this end, observe that the motion is
generated from the Hamiltonian:
\begin{equation}
H=\frac{{\bf p}^{2}}{2m}+V+Q,
\end{equation}
which is a quantum Hamilton-Jacobi equation written in terms of ${\bf p}\equiv\nabla S$.
The particle velocity is given by de Broglie's guidance equation:
\begin{align}
{\bf \dot{x}}= & \frac{\nabla_{{\bf x}}S}{m}=\frac{{\bf p}}{m}=\nabla_{{\bf {\bf p}}}H,\label{eq:G16}
\end{align}
while the particle momentum is given by the Bohmian equation of motion:
\begin{align}
\dot{{\bf {\bf p}}}= & -\nabla_{{\bf x}}V-\nabla_{{\bf x}}Q\label{eq:G17}\\
= & -\nabla_{{\bf {\bf x}}}H.
\end{align}
Now using these definitions, it is easy to prove that:
\begin{align}
 & \delta({\bf x}-{\bf X}_{t}^{a})\delta({\bf p}-{\bf P}_{t}^{a})\nabla_{{\bf x}}\cdot\dot{{\bf x}}+\delta({\bf x}-{\bf X}_{t}^{a})\delta({\bf p}-{\bf P}_{t}^{a})\nabla_{{\bf {\bf p}}}\cdot\dot{{\bf p}}\nonumber \\
= & \delta({\bf x}-{\bf X}_{t}^{a})\delta({\bf p}-{\bf P}_{t}^{a})\left(\nabla_{{\bf x}}\cdot\dot{{\bf x}}+\nabla_{{\bf {\bf p}}}\cdot\dot{{\bf p}}\right)\\
= & \delta({\bf x}-{\bf X}_{t}^{a})\delta({\bf p}-{\bf P}_{t}^{a})\left(\nabla_{{\bf x}}\cdot\nabla_{{\bf {\bf p}}}H-\nabla_{{\bf {\bf p}}}\cdot\nabla_{{\bf {\bf x}}}H\right)\\
= & 0,
\end{align}
since the derivative operators commute $\nabla_{{\bf x}}\cdot\nabla_{{\bf {\bf p}}}=\nabla_{{\bf {\bf p}}}\cdot\nabla_{{\bf x}}$.
Therefore we have an additional piece that can be added to {[}Eq.
\ref{eq:G.15}{]}:
\begin{equation}
\lim_{N\rightarrow\infty}\frac{1}{N}\sum_{a=1}^{N}\delta({\bf x}-{\bf X}_{t}^{a})\delta({\bf p}-{\bf P}_{t}^{a})\left(\nabla_{{\bf x}}\cdot\dot{{\bf x}}+\nabla_{{\bf {\bf p}}}\cdot\dot{{\bf p}}\right)=0,
\end{equation}
or equivalently:
\begin{equation}
\rho\nabla_{{\bf x}}\cdot\dot{{\bf x}}+\rho\nabla_{{\bf {\bf p}}}\cdot\dot{{\bf p}}=0.
\end{equation}
Adding this to {[}Eq. \ref{eq:G.15}{]} gives the continuity equation
in phase space:
\begin{align}
0= & \frac{\partial\rho}{\partial t}+\dot{{\bf x}}\cdot\nabla_{{\bf x}}\rho+\rho\nabla_{{\bf x}}\cdot\dot{{\bf x}}+\dot{{\bf p}}\cdot\nabla_{{\bf p}}\rho+\rho\nabla_{{\bf {\bf p}}}\cdot\dot{{\bf p}}\\
= & \frac{\partial\rho}{\partial t}+\nabla_{{\bf x}}\cdot\left(\rho\dot{{\bf x}}\right)+\nabla_{{\bf p}}\cdot\left(\rho\dot{{\bf p}}\right).
\end{align}
The continuity equation in phase space can be integrated over the
momentum coordinates to give the continuity equation in position space,
which follows from applying the divergence theorem i.e.:
\begin{equation}
\int\nabla_{{\bf p}}\cdot\left(\rho\dot{{\bf p}}\right)d{\bf p}=0,
\end{equation}
therefore:
\begin{align}
0= & \frac{\partial\int\rho({\bf x},{\bf p},t)d{\bf p}}{\partial t}+\nabla_{{\bf x}}\cdot\left(\int\rho({\bf x},{\bf p},t)\dot{{\bf x}}d{\bf p}\right)+\int\nabla_{{\bf p}}\cdot\left(\rho\dot{{\bf p}}\right)d{\bf p}\\
= & \frac{\partial\rho({\bf x},t)}{\partial t}+\nabla_{{\bf x}}\cdot\left(\rho({\bf x},t)\dot{{\bf x}}\right).
\end{align}
As we chosen $\dot{{\bf x}}=\frac{\nabla_{{\bf x}}S}{m}$ in {[}Eq.
\ref{eq:G16}{]} this equation becomes: 
\begin{align}
\frac{\partial\rho({\bf x},t)}{\partial t}+\nabla_{{\bf x}}\cdot\left(\rho({\bf x},t)\frac{\nabla_{{\bf x}}S}{m}\right)= & 0.
\end{align}
The continuity equation for the particle distribution {[}Eq. \ref{eq:H3}{]}
is identical to the continuity equation of quantum mechanics, therefore
the choices of $\dot{{\bf x}}=\frac{\nabla_{{\bf x}}S}{m}$ {[}Eq.
\ref{eq:G16}{]} and $\dot{{\bf {\bf p}}}=-\nabla_{{\bf x}}V-\nabla_{{\bf x}}Q$
{[}Eq. \ref{eq:G17}{]}, which are equations of the de Broglie-Bohm
interpretation, ensure the equivariance property. The particle distribution
will match the quantum mechanical probability density for all subsequent
times if prepared with the same distribution initially, due to the
uniqueness of solutions to the continuity equations. 
\end{document}